\documentclass[10pt]{article}

\usepackage{graphicx}

\usepackage{lineno,hyperref}
\usepackage{amssymb,amsmath,latexsym}
\usepackage{mathrsfs}
\usepackage{tipa}
\usepackage{amsfonts}
\usepackage{graphicx}
\usepackage{subfigure}
\usepackage[title]{appendix}
\usepackage{color}
\usepackage{multirow}
\usepackage{amsbsy}
\usepackage{longtable}
\usepackage{floatrow}
\usepackage{subfigure}
\usepackage{indentfirst}
\usepackage{paralist}
\newtheorem{theorem}{Theorem}
\newtheorem{definition}{Definition}
\newtheorem{proposition}{Proposition}

\newtheorem{cor}{Corollary}
\usepackage[font=small]{caption}
\usepackage{algorithm}
\usepackage{algorithmic}
\modulolinenumbers[5]
\usepackage{xcolor}
\usepackage{hyperref} 
\usepackage[paperwidth=8in,paperheight=10in, margin=.875in]{geometry}

\AtBeginDocument{\hypersetup{pdfborder={0 0 1}}}

\begin{document}

\title{\bf Minimization of the $q$-ratio sparsity with $1<q\leq \infty$ for signal recovery}

	\author {Zhiyong Zhou\footnote{Corresponding author, zhiyongzhou@zucc.edu.cn. This work was supported by the Swedish Research Council grant (Reg.No. 340-2013-5342).} \\              Department of Statistics, Zhejiang University City College, \\ Hangzhou,
              310015, China    
              \\  \\
              Jun Yu\\ 
              
		Department of Mathematics and Mathematical Statistics, Ume{\aa} University, \\Ume{\aa},
		901 87, Sweden}
	\maketitle
	\date{}
	\noindent

\maketitle

\begin{abstract}
In this paper, we propose a general scale invariant approach for sparse signal recovery via the minimization of the $q$-ratio sparsity. When $1<q\leq \infty$, both the theoretical analysis based on $q$-ratio constrained minimal singular values (CMSV) and the practical algorithms via nonlinear fractional programming are presented. Numerical experiments are conducted to demonstrate the advantageous performance of the proposed approaches over the state-of-the-art sparse recovery methods.

\end{abstract}

\noindent
{\bf Keywords}: Compressive sensing, $q$-ratio sparsity, $q$-ratio CMSV, nonlinear fractional programming, convex-concave procedure\\

\noindent
{\bf Mathematics Subject Classification (2010)}: MSC 94A12, MSC 94A20.

\section{Introduction}

Over the past decade, an extensive literature on Compressive Sensing has been developed, see the monographs \cite{eldar2012compressed,foucart2013mathematical} for a comprehensive view. Compressive Sensing aims to find the sparsest solution $x\in\mathbb{R}^N$ from few noisy linear measurements $y=Ax+\varepsilon\in\mathbb{R}^{m}$ where $A\in\mathbb{R}^{m\times N}$ with $m\ll N$, and $\lVert \varepsilon\rVert_2\leq \eta$, which can be formulated as solving a constrained $\ell_0$-minimization problem: \begin{align}
\min\limits_{z\in\mathbb{R}^N} \lVert z\rVert_0\quad \text{subject to \quad $\lVert Az-y\rVert_2\leq \eta $}. \label{l0-mini}
\end{align}
Unfortunately, it's a combinatorial problem which is known to be computationally NP-hard to solve \cite{natarajan1995sparse}. Instead, a widely used solver is the following constrained $\ell_1$-minimization problem \cite{donoho2006compressed}:
\begin{align}
\min\limits_{z\in\mathbb{R}^N} \lVert z\rVert_1\quad \text{subject to \quad $\lVert Az-y\rVert_2\leq \eta $},
\end{align}
which acts as a convex relaxation of $\ell_0$-minimization. 

Besides, a variety of non-convex recovery methods have been proposed to enhance sparsity, including $\ell_p$ ($0<p<1$) \cite{chartrand2007exact,chartrand2008iteratively,chartrand2008restricted,foucart2009sparsest,xu2012}, $\ell_1-\ell_2$ \cite{lou2018fast,lou2015computing,yin2015minimization}, transformed $\ell_1$ (TL1) \cite{zhang2018minimization}, SCAD\cite{fan2001variable}, MCP\cite{zhang2010nearly}, and $\ell_1/\ell_2$ \cite{rahimi2018scale,wang2020accelerated}, to name a few. These non-convex methods result in the difficulties of theoretical analysis and computational algorithms, but do lead to better recovery performances compared to the convex $\ell_1$-minimization in certain contexts. Among them, $\ell_p$ gives superior results for incoherent measurement matrices, while $\ell_1-\ell_2$ and $\ell_1/\ell_2$ are better choices for highly coherent measurement matrices. TL1 is a robust choice no matter whether the measurement matrix is coherent or not. 

Regarding the $\ell_1/\ell_2$ method, there are few literatures concerning on it due to its complex structure. In fact, $\ell_1/\ell_2$ is neither convex nor concave, and it's not even globally continuous. Some theoretical analyses have been done when it's restricted to non-negative signals\cite{esser2013method,yin2014ratio}. Very recently, a few new attempts have been made. \cite{rahimi2018scale} gave some local optimality results and proposed to solve it based on the Alternating Direction Method of Multipliers (ADMM)\cite{boyd2011distributed}. Some accelerated schemes were used for the $\ell_1/\ell_2$ minimization in \cite{wang2020accelerated}. \cite{wang2020limited} adopted the $\ell_1/\ell_2$ minimization in computed tomography (CT) reconstruction. \cite{xu2020analysis} investigated exact recovery conditions as well as the stability of the $\ell_1/\ell_2$ method, and provided the conditions under which $\ell_0$ minimization is equivalent to $\ell_1/\ell_2$ minimization. \cite{petrosyan2019reconstruction} showed that the $\ell_1/\ell_2$ method outperforms the $\ell_1$ minimization in jointly sparse vectors reconstruction problems and can be effectively solved via manifold optimization methods. Meanwhile, \cite{boct2020extrapolated} proposed a proximal subgradient algorithm with extrapolations for solving nonconvex and nonsmooth fractional programmings which encompass the $\ell_1/\ell_2$ minization problem. 

Inspired by the fact that $\lVert z\rVert_1^2/\lVert z\rVert_2^2$ is a special case of the $q$-ratio sparsity measure $s_q(z)=\left(\frac{\lVert z\rVert_1}{\lVert z\rVert_q}\right)^{\frac{q}{q-1}}$ with $q=2$, we propose in this paper a more general scale invariant approach for sparse signal recovery via minimizing the $q$-ratio sparsity measure $s_q(\cdot)$. We aim to theoretically and numerically investigate the minimization of the $q$-ratio sparsity. The main contribution of the present paper is four folds: (1) We gave a further study on the properties of $q$-ratio sparsity and illustrate them with examples. (2) We proposed the minimization of $q$-ratio sparsity for sparse signal recovery, which encompasses the well-known $\ell_0$-minimization, $\ell_1/\ell_2$ and $\ell_1/\ell_{\infty}$ methods. (3) We gave a verifiable sufficient condition for the exact sparse recovery, and introduced $q$-ratio constrained minimal singular values (CMSV) and derived concise bounds on both $\ell_q$ norm and $\ell_1$ norm of the reconstruction error for the $\ell_1/\ell_q$ with $q\in(1,\infty]$ in terms of $q$-ratio CMSV. We established the corresponding stable and robust recovery results involving both sparsity defect and measurement error. To the best of our knowledge, in the study of this kind of non-convex methods we are the first to establish the results for the compressible (not exactly sparse) case since all the literature mentioned above merely considered the exactly sparse case. (4) We presented efficient algorithms to solve the proposed methods via nonlinear fractional programming and conducted various numerical experiments to illustrate their superior performance.

The paper is organized as follows. In Section 2, we present the definition of $q$-ratio sparsity and some further study on its properties. In Section 3, we propose the sparse signal recovery methodology via the minimization of $q$-ratio sparsity. In Section 4, we provide a verifiable sufficient condition for the exact sparse recovery and derive the reconstruction error bounds based on $q$-ratio CMSV for the proposed method in the case of $1<q\leq \infty$. In Section 5, we design algorithms to solve the problem. Section 6 contains the numerical experiments. Finally, conclusions and future works are included in Section 7. 

Throughout the paper, we denote vectors by lower case letters e.g., $x$, and matrices by upper case letters e.g., $A$. Vectors are columns by defaults. $x^T$ denotes the transpose of $x$, while the notation $x_i$ denotes the $i$-th component of $x$. We introduce the notations $[N]$ for the set $\{1,2,\cdots,N\}$ and $|S|$ for the cardinality of a set $S$. Furthermore, we write $S^c$ for the complement $[N]\setminus S$ of a set $S$ in $[N]$. The support of a vector $x\in\mathbb{R}^N$ is the index set of its nonzero entries, i.e., $\mathrm{supp}(x):= \{i \in [N]: x_i\neq 0\}$. For any vector $x\in\mathbb{R}^N$, we denote by $\lVert x\rVert_0=\sum_{i=1}^N 1_{\{x_i\neq 0\}}=|\mathrm{supp}(x)|$ and we say $x$ is $k$-sparse if at most $k$ of its entries are nonzero, i.e., if $\lVert x\rVert_0\leq k$. The $\ell_q$-norm $\lVert x\rVert_q=(\sum_{i=1}^N |x_i|^q)^{1/q}$ for any $q\in(0,\infty)$, while $\lVert x\rVert_\infty=\max_{1\leq i\leq N} |x_i|$. For a vector $x\in\mathbb{R}^N$ and a set $S\subseteq [N]$, we denote by $x_S$ the vector which coincides with $x$ on the indices in $S$ and is extended to zero outside $S$. In addition, for any matrix $A\in\mathbb{R}^{m\times N}$, we denote the kernel of $A$ by $\mathrm{ker}(A)=\{x\in\mathbb{R}^N|Ax=0\}$.

\section{$q$-ratio sparsity}

In order to be self-contained, we first give the full definition of the $q$-ratio sparsity and then present some further study on its properties, together with some illustrative examples.

\begin{definition}(\cite{lopes2016unknown,zhou2018sparse})
	For any non-zero $z\in\mathbb{R}^N$ and non-negative $q\notin\{0,1,\infty\}$, the $q$-ratio sparsity level of $z$ is defined as \begin{align}
	s_{q}(z)=\exp(H_q(\pi(z)))=\left(\frac{\lVert z\rVert_1}{\lVert z\rVert_q}\right)^{\frac{q}{q-1}}, \label{sparsity_def}
	\end{align}
	where $H_q(\cdot)$ is the R\'{e}nyi entropy of order $q\in[0,\infty]$ \cite{plan2013one,vershynin2015estimation}. When $q\neq \{0,1,\infty\}$, the equality (\ref{sparsity_def}) holds since the R\'{e}nyi entropy is given by $H_q(\pi(z))=\frac{1}{1-q}\ln (\sum_{i=1}^N \pi_i(z)^q)$, while the cases of $q\in\{0,1,\infty\}$ are evaluated as limits: 
	$s_0(z)=\lim\limits_{q\rightarrow 0} s_q(z)=\lVert z\rVert_0$, $s_1(z)=\lim\limits_{q\rightarrow 1} s_q(z)=\exp(H_1(\pi(z)))$, $s_\infty(z)=\lim\limits_{q\rightarrow \infty} s_q(z)=\frac{\lVert z\rVert_1}{\lVert z \rVert_\infty}$.
	Here $\pi(z)\in\mathbb{R}^N$ with entries $\pi_i(z)=|z_i|/\lVert z\rVert_1$ and $H_1$ is the ordinary Shannon entropy $H_1(\pi(z))=-\sum_{i=1}^N \pi_i(z)\ln \pi_i(z)$.
\end{definition}

\begin{figure}[htbp]
	\centering
	\includegraphics[width=\textwidth,height=0.4\textheight]{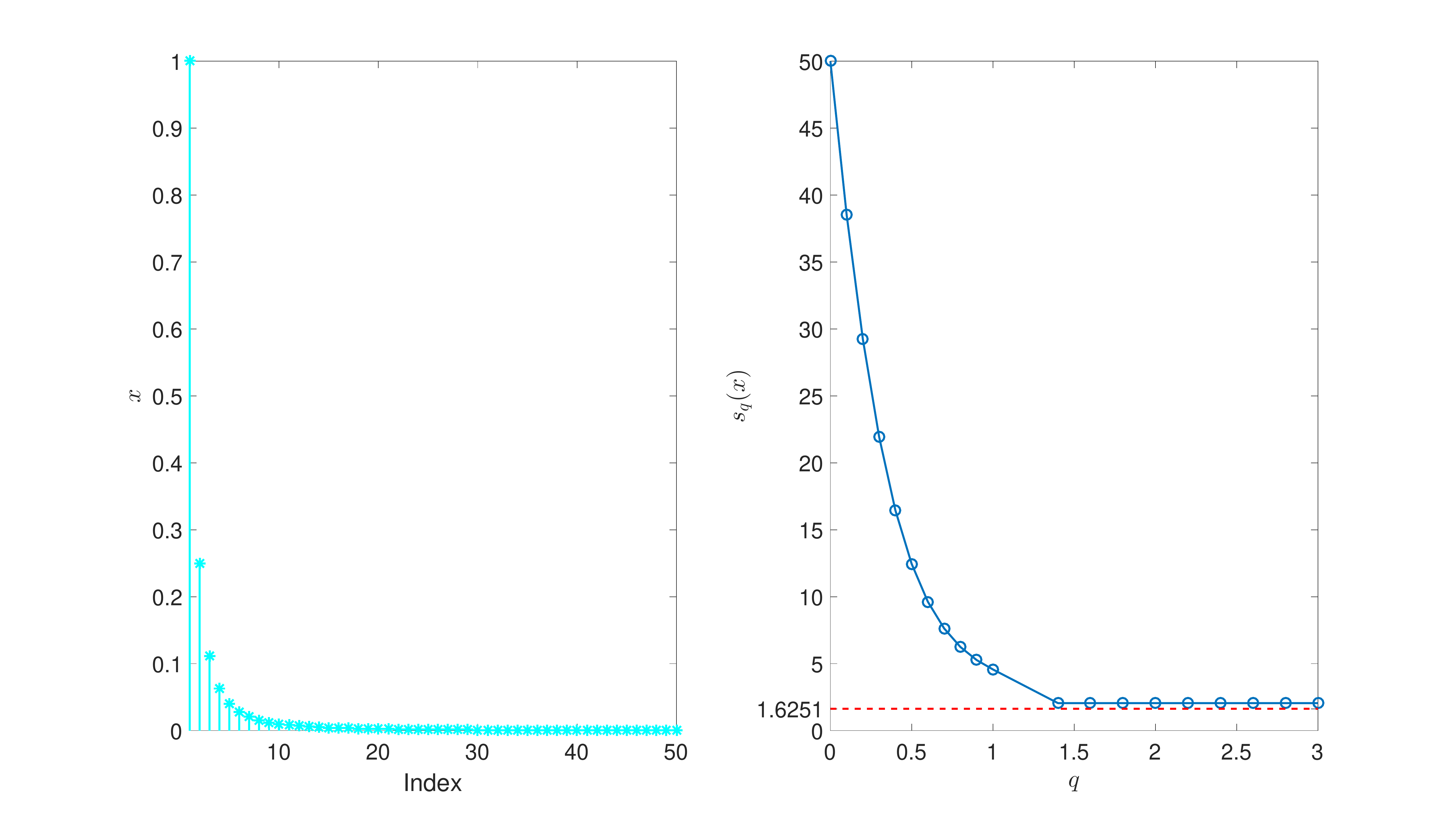}
	\caption{$q$-ratio sparsity levels $s_q(x)$ for a compressible signal $x$ while varying $q$. Here $x\in\mathbb{R}^N$ is generated with its entries decay as $i^{-2}$ where $i\in\{1,2,\cdots,50\}$. In the right panel, the dashed red line corresponds to the sparsity level $s_\infty(x)=\frac{\lVert x\rVert_1}{\lVert x\rVert_{\infty}}=1.6251$.} \label{fig:1}
\end{figure}

As an entropy-based sparsity measure, $s_q(\cdot)$ possesses important properties such as continuity, scale-invariance, non-increasing with respect to $q$ and range equal to $[1,N]$ (i.e., for any $q'\geq q \geq 0$, $1\leq \frac{\lVert z\rVert_{1}}{\lVert z\rVert_{\infty}}=s_\infty(z)\leq s_{q'}(z)\leq s_{q}(z)\leq s_{0}(z)=\lVert z\rVert_{0}\leq N$), see \cite{lopes2016unknown,zhou2018sparse} for details. To illustrate this sparsity measure, we show the corresponding $q$-ratio sparsity levels for a compressible signal of length 50 generated with its entries decay as $i^{-2}$ with $i\in \{1,2,\cdots,50\}$ in Figure \ref{fig:1}. As is shown, for a compressible signal with very small but non-zero entries, the $q$-ratio sparsity with a moderate large $q$ provided a better sparsity measure than the traditional $\ell_0$ norm. Another simple fact about the $q$-ratio sparsity is that, $s_q(u)<s_q(v)$ for some $q$, does not imply $\lVert u\rVert_0<\lVert v\rVert_0$. For instance, let $u=(10,0,1,0.1,0)^T$ and $v=(1,1,0,0)^T$, although we have $s_2(u)=\left(\frac{\lVert u\rVert_1}{\lVert u\rVert_2}\right)^{2}=1.0402<s_2(v)=\left(\frac{\lVert v\rVert_1}{\lVert v\rVert_2}\right)^{2}=2$, but apparently $\lVert u\rVert_0=3>\lVert v\rVert_0=2$. 

\begin{figure}[htbp]
	\centering
	\includegraphics[width=\textwidth,height=0.4\textheight]{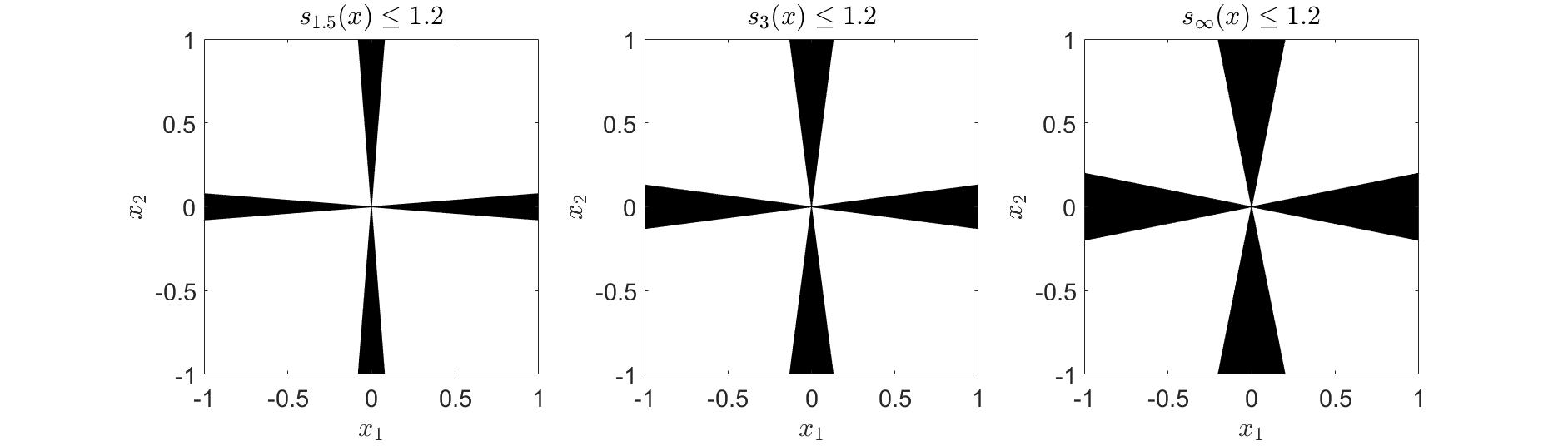}
	\caption{$q$-ratio sparsity level sets $S_{q,1.2}=\{x\in\mathbb{R}^2: s_q(x)\leq 1.2\}$ with $q=1.5,3,\infty$.} \label{fig:2}
\end{figure}

Let $\Sigma_k:=\{z\in\mathbb{R}^N: \lVert z\rVert_0\leq k\}$ and $S_{q,k}:=\{z\in\mathbb{R}^N: s_q(z)\leq k\}$ be the sparsity level set and the $q$-ratio sparsity level set, respectively. The following two propositions show that the $q$-ratio sparsity level set $S_{q,k}$ contains both the exactly $k$-sparse and the compressible vectors (by letting $s=N$ in Proposition 2, the vector $z$ considered there is not exactly sparse but can be well approximated by a $k$-sparse vector if $\xi$ is sufficiently small). They can be viewed as extensions of Proposition 1 and Proposition 2 in \cite{zc}, which only considered the special case of $q=2$. An illustrative example for the sets $S_{q,1.2}$ with $q=1.5,2,\infty$ in $\mathbb{R}^2$ is presented in Figure \ref{fig:2}. 

\begin{proposition}
	For any $0\leq q_1\leq q_2\leq \infty$, we have $S_{q_1,k}\subseteq S_{q_2,k}$. In particular, for any $q\in [0,\infty]$, it holds that $\Sigma_k\subseteq S_{q,k}$.
\end{proposition}

\noindent
{\bf Proof.} This follows immediately from the non-increasing property of $q$-ratio sparsity level with respect to $q$, i.e., for any $0\leq q_1\leq q_2\leq \infty$ and $z\in S_{q_1,k}$, we have $s_{q_2}(z)\leq s_{q_1}(z)\leq k$, hence $z\in S_{q_2,k}$. The second half statement follows by letting $q_1=0$ and $q_2=q$.

\begin{proposition}
	Consider an $s$-sparse vector $z\in\mathbb{R}^N$ with $s>k$. For any $1<q\leq \infty$, assume that $I\subset \mathrm{supp}(z)$, $|I|<k$ and $\lVert z_{I^c}\rVert_q\leq \xi$. We have $z\in S_{q,k}$ for some sufficiently small positive $\xi$. 
\end{proposition}

\noindent
{\bf Proof.} We denote $t_1=|I|$ and $t_2=|I^c|$. Then for any $q\in (1,\infty]$, we have \begin{align*}
\lVert z\rVert_1&=\lVert z_I\rVert_1+\lVert z_{I^c}\rVert_1 \\
&\leq t_1^{1-1/q}\lVert z_I\rVert_q+t_2^{1-1/q}\lVert z_{I^c}\rVert_q \\
&\leq t_1^{1-1/q}\lVert z_I\rVert_q+t_2^{1-1/q}\xi.
\end{align*}
Since $(k^{1-1/q}-t_1^{1-1/q})\lVert z_I\rVert_q=(k^{1-1/q}-|I|^{1-1/q})\lVert z_I\rVert_q>0$, we can choose a sufficient small $\xi$ such that $\xi\leq t_2^{1/q-1}(k^{1-1/q}-t_1^{1-1/q})\lVert z_I\rVert_q$. Thus, we obtain $\lVert z\rVert_1\leq k^{1-1/q}\lVert z_I\rVert_q\leq  k^{1-1/q}\lVert z\rVert_q$ which leads to $s_q(z)=\left(\frac{\lVert z\rVert_1}{\lVert z\rVert_q}\right)^{\frac{q}{q-1}}\leq k$, and hence $z\in S_{q,k}$. \\

\section{Methodology}

Based on the $q$-ratio sparsity $s_q(\cdot)$, we here consider the following non-convex minimization problem for sparse signal recovery: \begin{align}
\min\limits_{z\in\mathbb{R}^N} s_q(z)\quad \text{subject to \quad $\lVert y-Az\rVert_2\leq \eta$}, \label{sparsity_min}
\end{align}
where $y=Ax+\varepsilon$ with $\lVert \varepsilon\rVert_2\leq \eta$, and some $q\in[0,\infty]$ is pre-given. Obviously, when $q\rightarrow 0$, the problem approaches the $\ell_0$-minimization problem (\ref{l0-mini}) as $s_q(z)$ approaches $s_0(z)=\lVert z\rVert_0$.

\begin{figure}[htbp]
	\centering
	\includegraphics[width=\textwidth,height=0.4\textheight]{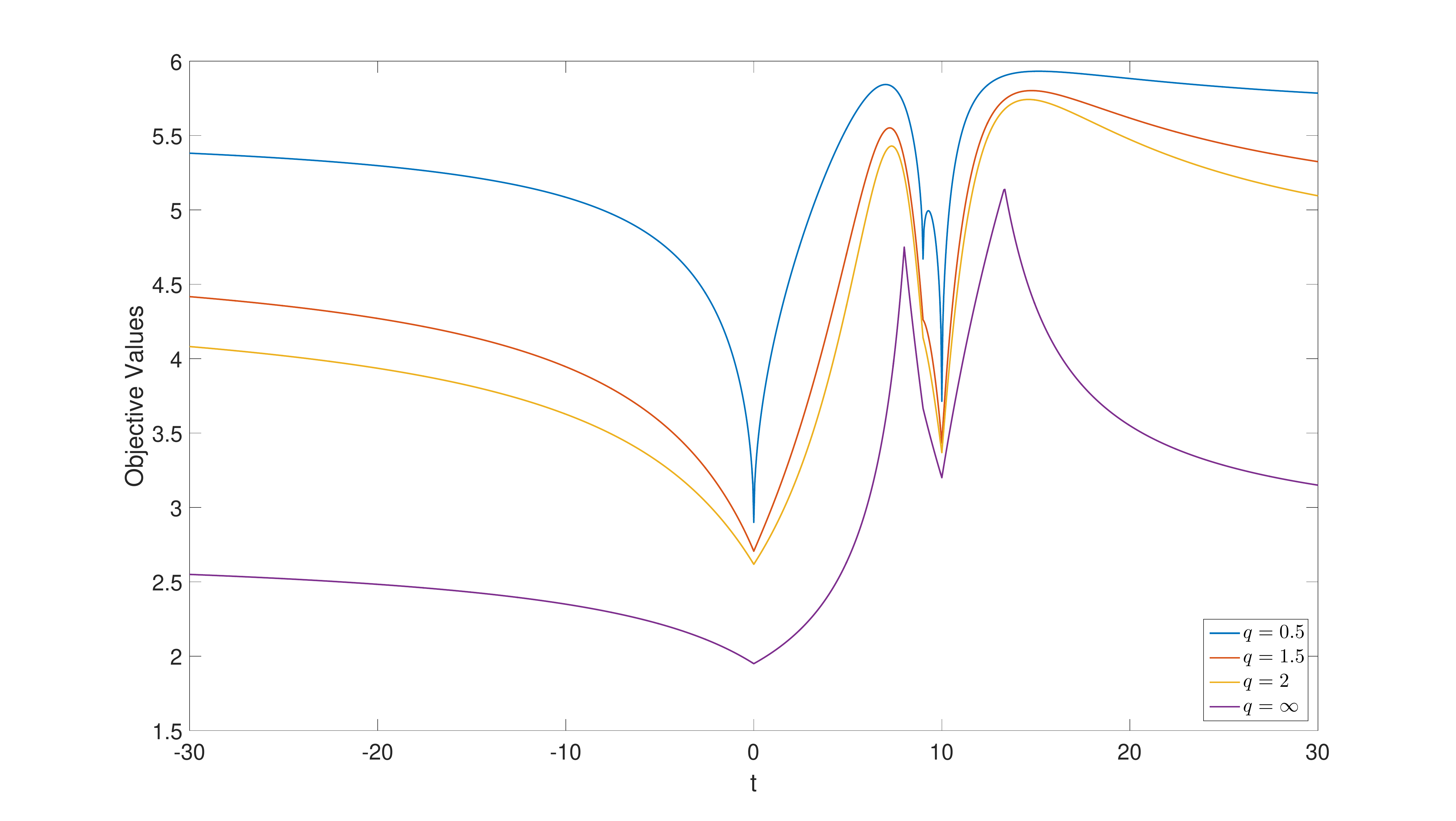} 
	\caption{The objective functions of a toy example used to illustrate that minimizing $q$-ratio sparsity $s_q(\cdot)$ can find $t=0$ as the global minimizer.} \label{toy_example}
\end{figure}

To illustrate the sparsity promoting ability of the problem (\ref{sparsity_min}), we revisit a toy example which was discussed in \cite{rahimi2018scale}. Specifically, let the measurement matrix $$A=\begin{pmatrix}
1& -1&  0& 0&  0&0 \\
1&  0& -1& 0&  0&0 \\
0& 1& 1& 1& 0& 0 \\
2& 2& 0& 0& 1& 0 \\
1& 1& 0& 0& 0& -1
\end{pmatrix}
\in\mathbb{R}^{5\times 6},$$ and the measurement vector $y=(0,0,20,40,18)^T\in\mathbb{R}^5$. Then, any solution of $Az=y$ has the form of $z=(t,t,t,20-2t,40-4t,2(t-9))^T$ for some $t\in\mathbb{R}$. And it's easy to notice that the sparsest solution occurs at $t=0$, where its sparsity is 3. Other local solutions include $t=10$ with sparsity being 4, and $t=9$ with sparsity being 5. As was shown in Figure 1 of \cite{rahimi2018scale}, among the methods discussed there (including $\ell_1$, $\ell_{0.5}$, TL1, $\ell_1-\ell_2$ and $\ell_1/\ell_2$), only TL1 and $\ell_1/\ell_2$ model (corresponding to our model with $q=2$) can find the global minimizer $t=0$. Moreover, according to Figure \ref{toy_example} presented here, other choices of $q$ are also able to find the global minimizer at $t=0$. Specifically, as we can see, for $q=1.5,2,\infty$, the objective functions have two local minimizers ($t=0$ and $t=10$), while it has three local minimizers ($t=0$, $t=9$ and $t=10$) when $q=0.5$. 

\begin{figure}[htbp]
	\centering
	\includegraphics[width=\textwidth,height=0.4\textheight]{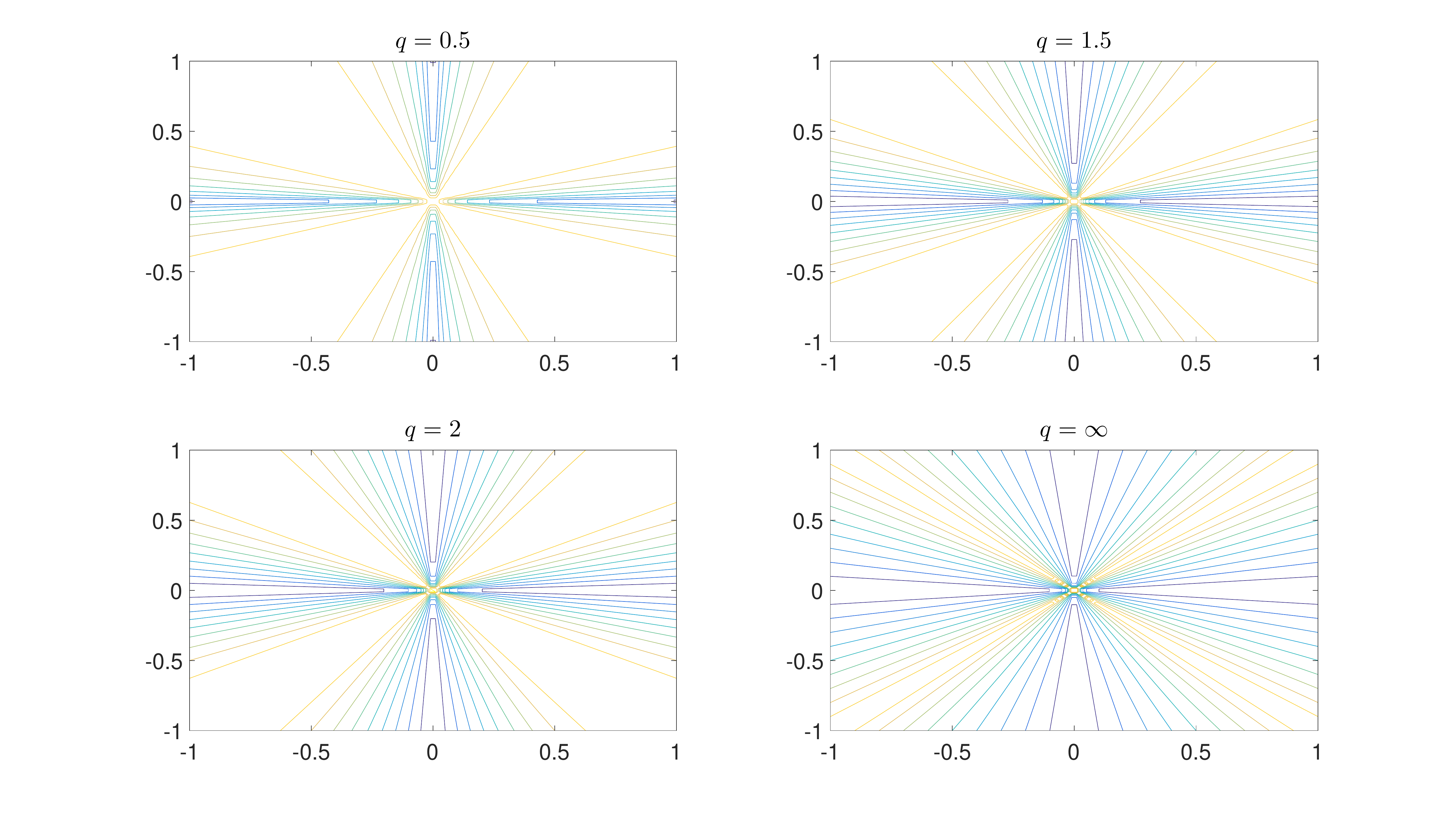} 
	\caption{The contour plots for $s_q(\cdot)$ with $q=0.5, 1.5, 2, \infty$. } \label{contour_plots}
\end{figure}

In addition, we present the contour plots for $s_q(\cdot)$ with different values of $q$ in Figure \ref{contour_plots}. As we can see, similar non-convex patterns arise while varying $q$. The fact that the level curves of $s_q(\cdot)$ approach the $x$ and $y$ axes as their values get small, also reflects their ability to promote sparsity. 

In Compressive Sensing, an incessant challenge is how to numerically solve the optimization problem. Consider our non-convex problem (\ref{sparsity_min}), the minimization structure of the case of $1<q\leq \infty$ is essentially different from the case of $0<q\leq 1$. Intuitively, it seems that it's much harder to solve the case of $0<q\leq 1$ than the case of $1<q\leq \infty$ because of the additional non-convexity of the $\ell_q$ norm with $0<q<1$. This can be observed in Figure \ref{toy_example} where it shows that minimizing $s_{0.5}(\cdot)$ processes more local minimizers compared to the cases of $q>1$.

As a preliminary exploration in this direction, in the following context we only focus on $1<q\leq \infty$, in which case solving (\ref{sparsity_min}) is equivalent to solve \begin{align}
\min\limits_{z\in\mathbb{R}^N} \frac{\lVert z\rVert_1}{\lVert z\rVert_q}\quad \text{subject to \quad $\lVert y-Az\rVert_2\leq \eta$}.\label{norm_ratio}
\end{align}
The case of $q=2$ without measurement errors (i.e., $\eta=0$) has been investigated in \cite{esser2013method,yin2014ratio} for non-negative signals, and very recently in \cite{rahimi2018scale,wang2020accelerated,xu2020analysis} for arbitrary signals. In the present paper, we extend it to the scenarios of any $1<q\leq \infty$ with measurement errors. The norm ratio $\lVert \cdot\rVert_1/\lVert \cdot\rVert_q$ can be regarded as normalized $\ell_1$ norm. In fact, it's also a special case of the sparsity measure called the $pq$-mean, with $p=1$ and some $q>1$, see \cite{hurley2009comparing,jia2018sparse} for details. 

Moreover, as discussed in Theorem 3.2 of \cite{pastor2015mathematics}, the case of $q=\infty$ (i.e., $\ell_1/\ell_\infty$-minimization) is almost equivalent to the ordinary $\ell_1$-minimization. Let $t=\frac{1}{|z_{\max}|}$, $u_i=tz_i$ and $v_i=\frac{1}{t}u_i$, then the proof goes as follows: 
\begin{align*}
\left\{
\begin{aligned}
&\min_{z}\frac{\lVert z\rVert_1}{\lVert z\rVert_\infty} \\
& \text{s.t.}\quad \lVert y-Az\rVert_2\leq \eta.
\end{aligned}
\right.=\left\{
\begin{aligned}
&\min_{z}\frac{\lVert z\rVert_1}{|z_{\max}|} \\
& \text{s.t.}\quad \lVert y-Az\rVert_2\leq \eta.
\end{aligned}
\right.
&=\left\{
\begin{aligned}
&\min_{u}\,\lVert u\rVert_1 \\
& \text{s.t.}\quad |u|\leq 1, t\geq 0, \lVert y-Au/t\rVert_2\leq \eta.
\end{aligned}
\right. \\
&=\left\{
\begin{aligned}
&\min_{v}\,\lVert v\rVert_1 \\
& \text{s.t.}\quad |v|\leq \frac{1}{t}, t\geq 0, \lVert y-Av\rVert_2\leq \eta.
\end{aligned}
\right.
\end{align*}

\section{Recovery analysis}

In this section, we study the global optimality results for the $\ell_1/\ell_q$ minimization with $q\in(1,\infty]$. We choose not to present the local optimality results based on null space property as given in \cite{rahimi2018scale,xu2020analysis}. We conjecture that the local optimality results for the $\ell_1/\ell_2$ minimization in \cite{rahimi2018scale,xu2020analysis} also hold for the $\ell_1/\ell_q$ minimization with $q\in(1,\infty]$ with some minor careful modifications.

We start with a sufficient condition for the exact sparse recovery using $\ell_1/\ell_q$ minimization with $q\in(1,\infty]$. For some pre-given $q\in(1,\infty]$, we consider the noiseless $\ell_1/\ell_q$ minimization problem:\begin{align}
\min\limits_{z\in\mathbb{R}^N} \frac{\lVert z\rVert_1}{\lVert z\rVert_q}\quad \text{subject to \quad $Az=Ax$}.\label{noiseless}
\end{align}
It's easy to see that when the true signal $x$ is $k$-sparse, the sufficient and necessary for the exact recovery of the problem (\ref{noiseless}) is given by the following null space property \cite{cohen2009compressed}: \begin{align}
\frac{\lVert x\rVert_1}{\lVert x\rVert_q}<\frac{\lVert x+h\rVert_1}{\lVert x+h\rVert_q},\quad \forall\,\, h\in \mathrm{ker}(A)\setminus \{0\}. \label{nsp}
\end{align}
Then, based on this null space property, we give the following sufficient condition that guarantees the uniform exact sparse recovery using the noiseless  $\ell_1/\ell_q$ problem (\ref{noiseless}).

\begin{proposition}
If $x$ is $k$-sparse, for some pre-given $q\in(1,\infty]$ such that $k$ is strictly less than \begin{align}
\inf\limits_{h\in \mathrm{ker}(A)\setminus \{0\}} 3^{\frac{q}{1-q}}s_q(h), \label{sufficient}
\end{align}
then the unique solution to the problem  (\ref{noiseless}) is the true signal $x$.
\end{proposition}

\noindent\\
{\bf Remark.} A similar sufficient condition was established in Section 4 of \cite{xu2020analysis} for the $\ell_p/\ell_2$ minimization problem with $0<p\leq 1$, while ours is for the $\ell_1/\ell_q$ minimization problem with $1<q\leq \infty$. As discussed in \cite{xu2020analysis}, this kind of sufficient condition is satisfied with high probability for sub-gaussian random matrix provided the number of its measurements $m$ behaves in a linearly in $k$ (up to a logarithmic factor). In addition, we notice that this kind of $q$-ratio sparsity based sufficient condition was also given for the noiseless $\ell_1$ minimization in Proposition 1 of \cite{zhou2018sparse}. The sufficient condition established there for the noiseless $\ell_1$-minimization is $k<\inf_{\{h\in \mathrm{ker}(A)\setminus \{0\}\}} 2^{\frac{q}{1-q}}s_q(h)$, which is slightly weaker than ours for the noiseless $\ell_1/\ell_q$ minimization. It is worth noting that the sufficient condition we obtained is verifiable for any pre-given measurement matrix. When $q=\infty$, the minimization problem (\ref{sufficient}) can be solved through $N$ linear programs with a polynomial time. In the cases of $1<q<\infty$, it’s very difficult to solve exactly, but can be solved approximately via a convex-concave procedure algorithm. Please see the Section 5.1 of \cite{zhou2018sparse} for details.

\noindent\\
{\bf Proof.} The proof follows the same route as the Proof of Theorem 4.1 in \cite{xu2020analysis}. It suffices to verify the null space property (\ref{nsp}). As $x$ is $k$-sparse, we can assume that $\mathrm{supp}(x)=S$ such that $|S|\leq k$. For any $h\in \mathrm{ker}(A)\setminus \{0\}$ and $q\in(1,\infty]$, since $\lVert x+h\rVert_1=\lVert x+h_S+h_{S^c}\rVert_1\geq \lVert x\rVert_1+\lVert h_{S^c}\rVert_1-\lVert h_S\rVert_1=\lVert x\rVert_1+\lVert h\rVert_1-2\lVert h_S\rVert_1$ and $\lVert x+h\rVert_q\leq \lVert x\rVert_q+\lVert h\rVert_q$, then it holds that\begin{align*}
\frac{\lVert x+h\rVert_1}{\lVert x+h\rVert_q}&\geq \frac{\lVert x\rVert_1+\lVert h\rVert_1-2\lVert h_S\rVert_1}{\lVert x\rVert_q+\lVert h\rVert_q} \\
&\geq \min\left\{\frac{\lVert x\rVert_1}{\lVert x\rVert_q},\frac{\lVert h\rVert_1-2\lVert h_S\rVert_1}{\lVert h\rVert_q}\right\}.
\end{align*}
When $k<3^{\frac{q}{1-q}}s_q(h)$, we have $\frac{\lVert h\rVert_1}{\lVert h\rVert_q}>3k^{1-1/q}>k^{1-1/q}+2\frac{\lVert h_S\rVert_1}{\lVert h\rVert_q}$, which implies that $\frac{\lVert h\rVert_1-2\lVert h_S\rVert_1}{\lVert h\rVert_q}>k^{1-1/q}\geq\frac{\lVert x\rVert_1}{\lVert x\rVert_q}$. Consequently, the null space property (\ref{nsp}) holds and the proof is complete. \\

In what follows, we study the stable and robust recovery analysis results for the minimization problem (\ref{norm_ratio}) involving both sparsity defect and measurement error. First, we present the definition of $q$-ratio constrained minimal singular values (CMSV), which is a computable quality measure for the measurement matrix. As an efficient theoretical analysis tool, $q$-ratio CMSV can be well used for convex sparse recovery algorithms such as the Basis Pursuit \cite{chen1998}, the Dantzig selector \cite{candes2007dantzig} and the Lasso \cite{tibshirani1996regression}, and also for non-convex weighted $\ell_r-\ell_1$ minimization with $r\in (0,1]$ \cite{zhou2019new}. More detailed arguments about $q$-ratio CMSV are referred \cite{tang2011performance,zhou2018q,zhou2018sparse}. 

\begin{definition}(\cite{zhou2018sparse})
	For any real number $s\in[1,N]$, $q\in(1,\infty]$ and matrix $A\in\mathbb{R}^{m\times N}$, the $q$-ratio constrained minimal singular value (CMSV) of $A$ is defined as \begin{align}
	\rho_{q,s}(A)=\min\limits_{z\neq 0,s_q(z)\leq s}\,\,\frac{\lVert Az\rVert_2}{\lVert z\rVert_q}. 
	\end{align} 
\end{definition}
\medskip

Next, we establish the uniform recovery analysis results for the problem (\ref{norm_ratio}) based on the $q$-ratio CMSV, which shows that the $q$-ratio CMSV also works pretty well in the recovery analysis for the non-convex $\ell_1/\ell_q$ minimization problem. We start with the following main result for the case that the true signal $x$ is exactly sparse. 
\newpage

\begin{theorem}(Exactly sparse recovery)
	Suppose $x$ is non-zero and $k$-sparse. For any $1<q\leq \infty$, if $\rho_{q,3^{\frac{q}{q-1}}k}(A)>0$, then the solution $\hat{x}$ to the problem (\ref{norm_ratio}) obeys \begin{align}
	\lVert \hat{x}-x\rVert_q &\leq \frac{2\eta}{\rho_{q,3^{\frac{q}{q-1}}k}(A)}, \\
	\lVert \hat{x}-x\rVert_1 &\leq  \frac{6 k^{1-1/q}\eta}{\rho_{q,3^{\frac{q}{q-1}}k}(A)}.
	\end{align}
\end{theorem}

\noindent\\
{\bf Remark.} As studied in \cite{zhou2018q,zhou2018sparse}, this sort of $q$-ratio CMSV based condition is fulfilled with high probability for subgaussian and some structured random matrix when the number of its measurements is reasonably large compared to the sparsity level $k$ (behaves in a linearly in $k$ up to a logarithmic factor). Moreover, for any pre-given measurement matrix $A$, its $q$-ratio CMSV can be computed approximately by using an interior point (IP) algorithm, see the Section 5.2 of \cite{zhou2018sparse} for detailed arguments. The stability analysis of noisy $\ell_1/\ell_2$ minimization has also been presented in \cite{xu2020analysis}, while the discussion there splits into two cases which is far less clean and straightforward than ours. 

\noindent\\
{\bf Proof.} Since $x$ is $k$-sparse, we have that $|S|\leq k$, where $S$ is the support of $x$. Let the residual $h=\hat{x}-x$. Since $\hat{x}=x+h$ is the minimum among all $z$ satisfying the constraint of (\ref{norm_ratio}), we have \begin{align*}
\frac{\lVert x+h\rVert_1}{\lVert x+h\rVert_q}\leq \frac{\lVert x\rVert_1}{\lVert x\rVert_q},
\end{align*}
which implies that \begin{align}
\lVert x+h\rVert_1\cdot\lVert x\rVert_q\leq \lVert x\rVert_1\cdot\lVert x+h\rVert_q. \label{ineq}
\end{align}
Moreover, \begin{align*}
\lVert x+h\rVert_1=\lVert x_S+h_S\rVert_1+\lVert x_{S^c}+h_{S^c}\rVert_1\geq \lVert x_S\rVert_1-\lVert h_S\rVert_1+\lVert h_{S^c}\rVert_1=\lVert x\rVert_1-\lVert h_S\rVert_1+\lVert h_{S^c}\rVert_1,
\end{align*}
and $\lVert x+h\rVert_q\leq \lVert x\rVert_q+\lVert h\rVert_q$. Therefore, it follows from (\ref{ineq}) that \begin{align*}
(\lVert x\rVert_1-\lVert h_S\rVert_1+\lVert h_{S^c}\rVert_1)\cdot\lVert x\rVert_q\leq \lVert x\rVert_1\cdot (\lVert x\rVert_q+\lVert h\rVert_q).
\end{align*}
Consequently, we obtain that \begin{align}
\lVert h_{S^c}\rVert_1\leq \lVert h_S\rVert_1+\frac{\lVert x\rVert_1}{\lVert x\rVert_q}\lVert h\rVert_q=\lVert h_S\rVert_1+s_q(x)^{1-1/q}\lVert h\rVert_q,
\end{align}
which leads to \begin{align}
\lVert h\rVert_1=\lVert h_S\rVert_1+\lVert h_{S^c}\rVert_1\leq 2\lVert h_S\rVert_1+s_q(x)^{1-1/q}\lVert h\rVert_q\leq (2k^{1-1/q}+s_q(x)^{1-1/q})\lVert h\rVert_q.
\end{align}
Thus, for any $1<q\leq \infty$, $s_q(h)=\left(\frac{\lVert h\rVert_1}{\lVert h\rVert_q}\right)^{\frac{q}{q-1}}\leq \left(2k^{1-1/q}+s_q(x)^{1-1/q}\right)^{\frac{q}{q-1}}\leq 3^{\frac{q}{q-1}} k$, by using the fact that $s_q(x)\leq \lVert x\rVert_0\leq k$. 

Since both $\hat{x}$ and $x$ satisfy the constraint $\lVert y-Az\rVert_2\leq \eta$, the triangle inequality implies \begin{align}
\lVert Ah\rVert_2=\lVert A(\hat{x}-x)\rVert_2\leq \lVert A\hat{x}-y\Vert_2+\lVert y-Ax\rVert_2\leq 2\eta. \label{Ah_bound}
\end{align}
Then, it follows from the definition of $q$-ratio CMSV and $s_q(h)\leq 3^{\frac{q}{q-1}}k$ that \begin{align*}
\rho_{q,3^{\frac{q}{q-1}}k}(A)\lVert h\rVert_q\leq \lVert Ah\rVert_2\leq 2\eta\Rightarrow \lVert h\rVert_q\leq \frac{2\eta}{\rho_{q,3^{\frac{q}{q-1}}k}(A)}.
\end{align*}
Meanwhile, $\lVert h\rVert_1\leq 3k^{1-1/q}\lVert h\rVert_q\Rightarrow \lVert h\rVert_1\leq \frac{6k^{1-1/q}\eta}{\rho_{q,3^{\frac{q}{q-1}}k}(A)}$, which completes the proof.  \\

As a by-product of Theorem 1, we have the following corollary immediately by letting $\eta=0$. The presented condition that $\rho_{q,3^{\frac{q}{q-1}}k}(A)>0$ is a bit stronger than the condition that $\rho_{q,2^{\frac{q}{q-1}}k}(A)>0$ given for the $\ell_1$-minimization in \cite{zhou2018sparse}.

\begin{cor}
For any $k$-sparse signal $x$ and any $q\in(1,\infty]$, if the condition $\rho_{q,3^{\frac{q}{q-1}}k}(A)>0$ holds, then the unique solution of (\ref{norm_ratio}) with $\eta=0$ is exactly the truth $x$. 
\end{cor}

\medskip
Furthermore, we extend the above result to the case when the true signal is allowed to be not exactly sparse, but is compressible, i.e., it can be well approximated by an exactly sparse signal. As far as we know, this part of result has not been touched before.

\begin{theorem}(Compressible recovery)
	Let the $\ell_1$-error of best $k$-term approximation of non-zero $x$ be $\sigma_{k,1}(x)=\inf \{\lVert x-z\rVert_1, z\in\mathbb{R}^N \text{is $k$-sparse}\}$, which is a function that measures how close $x$ is to being $k$-sparse. Denote $C_q(k,x)=(4k^{1-1/q}+s_q(x)^{1-1/q})^{\frac{q}{q-1}}$. For any $1<q\leq \infty$, if  $\rho_{q,C_q(k,x)}(A)>0$, then the solution $\hat{x}$ to the problem (\ref{norm_ratio}) obeys \begin{align}
	\lVert\hat{x}-x\rVert_{q}&\leq \frac{2\eta}{\rho_{q,C_q(k,x)}(A)}+k^{1/q-1}\sigma_{k,1}(x), \label{robust} \\
	\lVert\hat{x}-x\rVert_{1}&\leq \frac{(4k^{1-1/q}+2s_q(x)^{1-1/q})\eta}{\rho_{q,C_q(k,x)}(A)}+\left(4+(s_q(x)/k)^{1-1/q}\right)\sigma_{k,1}(x). \label{robustl1}
	\end{align}
\end{theorem}
\medskip
\noindent\\
{\bf Proof.} Assume that $S$ is the index set that contains the largest $k$ absolute entries of $x$ so that $\sigma_{k,1}(x)=\lVert x_{S^c}\rVert_1$ and let $h=\hat{x}-x$. Recall (\ref{ineq}) which holds here as well. Since \begin{align*}
\lVert x+h\rVert_1=\lVert x_S+h_S\rVert_1+\lVert x_{S^c}+h_{S^c}\rVert_1\geq \lVert x_S\rVert_1-\lVert h_S\rVert_1-\lVert x_{S^c}\rVert_1+\lVert h_{S^c}\rVert_1,
\end{align*} and $\lVert x+h\rVert_q\leq \lVert x\rVert_q+\lVert h\rVert_q$, it follows that \begin{align*}
(\lVert x_S\rVert_1-\lVert h_S\rVert_1-\lVert x_{S^c}\rVert_1+\lVert h_{S^c}\rVert_1)\cdot\lVert x\rVert_q\leq (\lVert x_S\rVert_1+\lVert x_{S^c}\rVert_1)\cdot \lVert x\rVert_q+\lVert x\rVert_1\cdot\lVert h\rVert_q.
\end{align*}
Consequently, we obtain that \begin{align}
\lVert h_{S^c}\rVert_1\leq \lVert h_S\rVert_1+2\lVert x_{S^c}\rVert_1+\frac{\lVert x\rVert_1}{\lVert x\rVert_q}\lVert h\rVert_q=\lVert h_S\rVert_1+2\lVert x_{S^c}\rVert_1+s_q(x)^{1-1/q}\lVert h\rVert_q,
\end{align}
which leads to \begin{align}
\lVert h\rVert_1=\lVert h_S\rVert_1+\lVert h_{S^c}\rVert_1&\leq 2\lVert h_S\rVert_1+2\lVert x_{S^c}\rVert_1+s_q(x)^{1-1/q}\lVert h\rVert_q  \nonumber\\
&\leq (2k^{1-1/q}+s_q(x)^{1-1/q})\lVert h\rVert_q+2\lVert x_{S^c}\rVert_1. \label{errorbound}
\end{align}

We assume that $h\neq 0$ and $\lVert h\rVert_q>\frac{2\eta}{\rho_{q,C_q(k,x)}(A)}$, otherwise (\ref{robust}) holds trivially. Since $\lVert Ah\rVert_2\leq 2\eta$, see (\ref{Ah_bound}), so we have $\lVert h\rVert_q>\frac{\lVert Ah\rVert_2}{\rho_{q,C_q(k,x)}(A)}$. Then it holds that \begin{align}
&\frac{\lVert Ah\rVert_2}{\lVert h\rVert_q}<{\rho_{q,C_q(k,x)}(A)}=\min\limits_{z\neq 0, s_q(z)\leq C_q(k,x)}\frac{\lVert Az\rVert_2}{\lVert z\rVert_q} \nonumber \\
&\quad \Rightarrow s_q(h)>C_q(k,x)\Rightarrow \lVert h\rVert_1>C_q(k,x)^{1-1/q}\lVert h\rVert_q=(4k^{1-1/q}+s_q(x)^{1-1/q})\lVert h\rVert_q.
\end{align}
Combining (\ref{errorbound}), we have $\lVert h\rVert_q<k^{1/q-1}\lVert x_{S^c}\rVert_1=k^{1/q-1}\sigma_{k,1}(x)$, which completes the proof of (\ref{robust}). The error $\ell_1$ norm bound (\ref{robustl1}) follows immediately from (\ref{robust}) and (\ref{errorbound}). 

\noindent\\
{\bf Remark.} For any compressible $x$, by adopting Proposition 2 with $z=x$, $s=N$, $I=S$, replacing $k$ by $k+c$ for any small constant $c>0$, and $\xi=\sigma_{k,1}(x)=\lVert x_{S^c}\rVert_1$, we have $s_q(x)\leq k+c$ as long as $\sigma_{k,1}(x)$ is sufficiently small. We can see that the reconstruction error bounds consist of two components, one is caused by the measurement error, while the other one is caused by the sparsity defect. And according to the proof procedure presented, we can sharpen the error bounds to be the maximum of these two components instead of their summation. \\

\begin{figure*}
	\centering
	\begin{subfigure}
		\centering
		\includegraphics[width=0.48\textwidth,height=0.3\textheight]{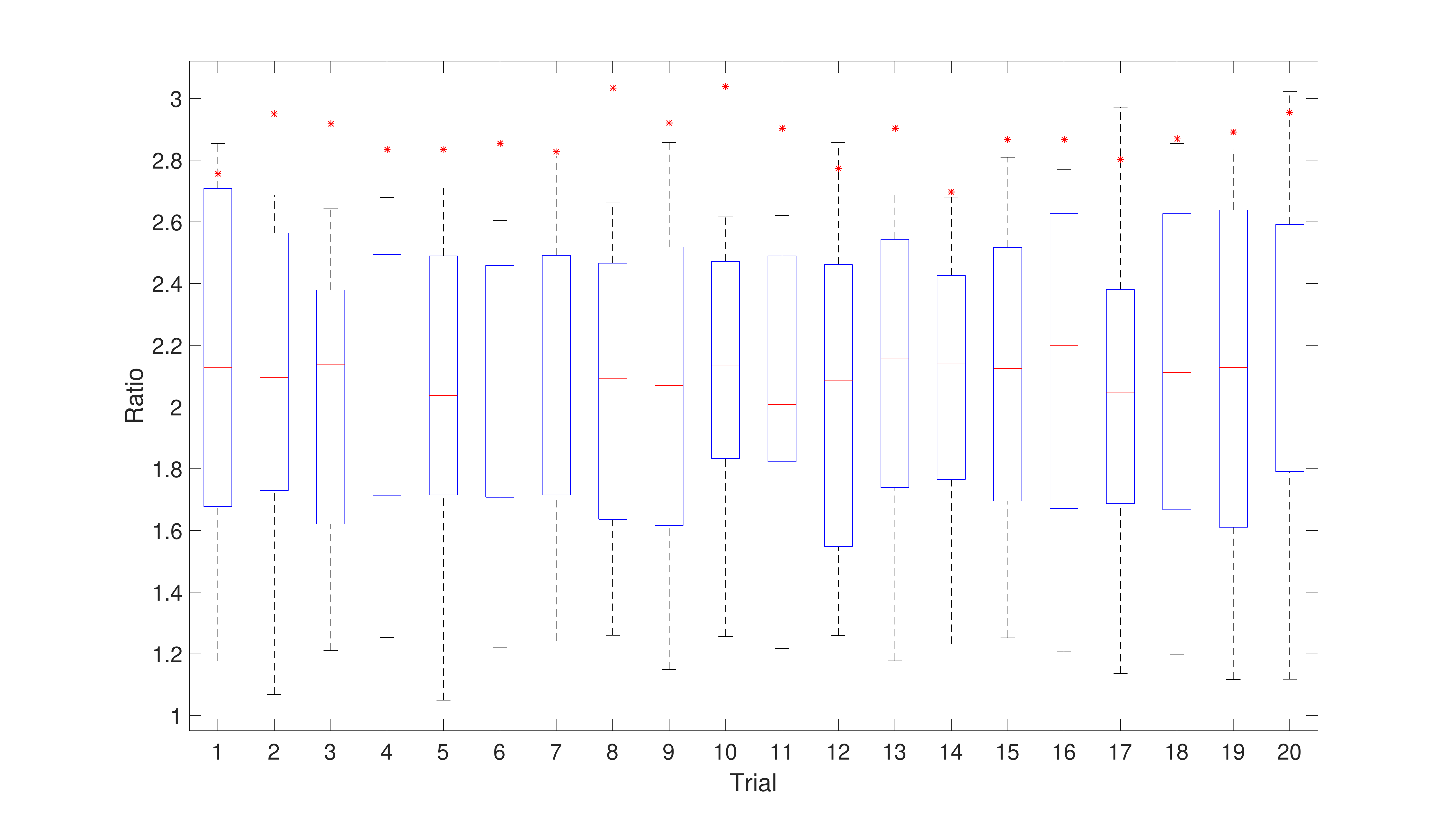}
	\end{subfigure}
	\begin{subfigure}
		\centering
		\includegraphics[width=0.48\textwidth,height=0.3\textheight]{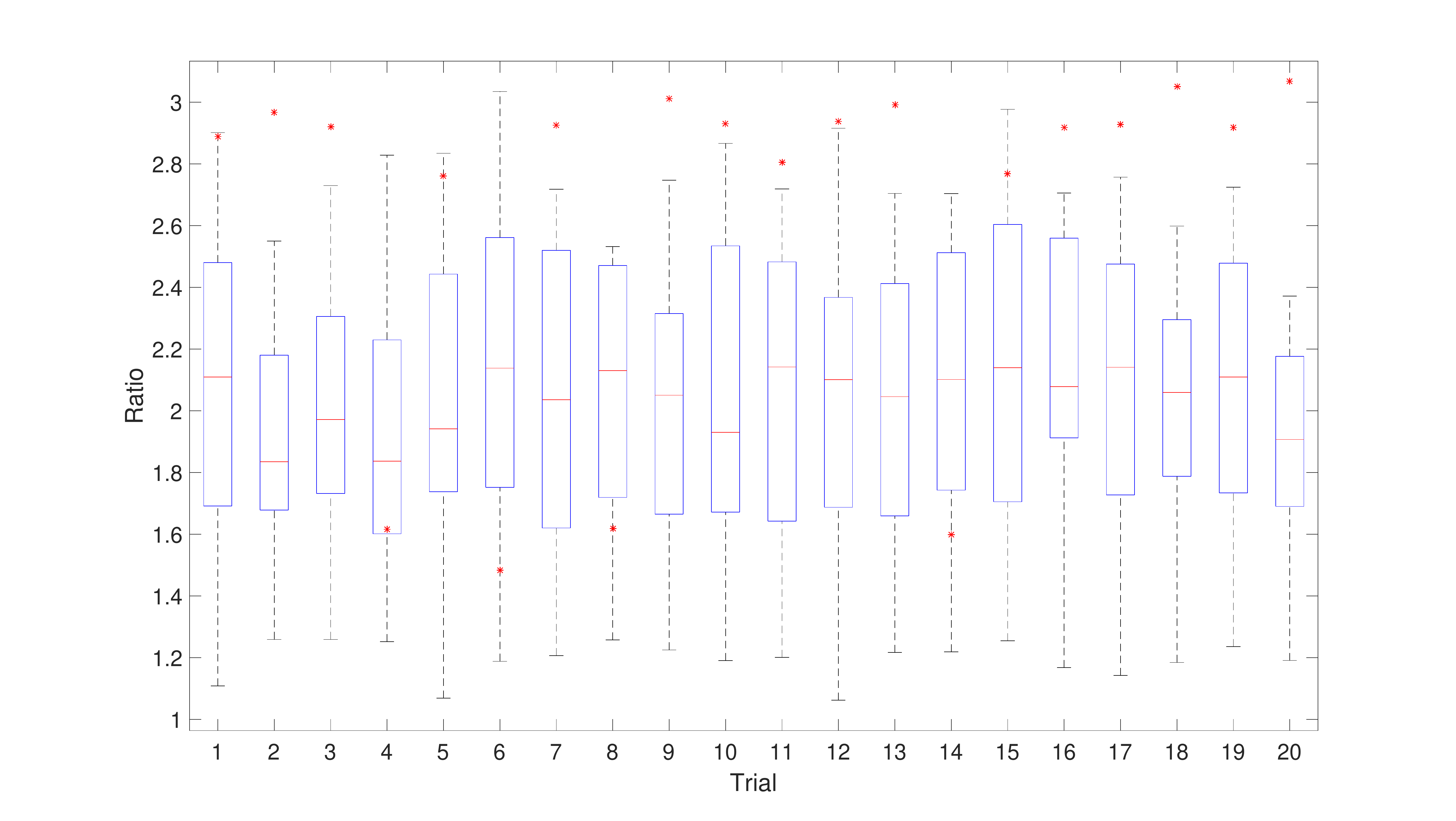}
	\end{subfigure}
	\caption{The ratios $\inf\limits_{z\in\mathbb{R}^N} \left\{ \frac{\lVert z\rVert_1}{\lVert z\rVert_q}\Big| Az=Ax\right\}$ (box-plot) and $\inf\limits_{z\in\mathbb{R}^N} \left\{ \frac{\lVert z\rVert_1}{\lVert z\rVert_q}\Big| z\in\mathrm{ker} (A)\setminus \{0\}\right\}$ (star-point) for DCT matrix, with $q=1.5$. Left panel ($F=2$), Right panel ($F=5$).} \label{ratio_comparison}
\end{figure*}

Moreover, a similar fact which acts as an extension of Proposition 1 in \cite{rahimi2018scale} is listed as follows. We skip the proof details since it follows from almost the same procedure as the proof in \cite{rahimi2018scale}. This proposition is empirically verified in Figure \ref{ratio_comparison}, where we calculate the ratio in $\inf\limits_{z\in\mathbb{R}^N} \left\{ \frac{\lVert z\rVert_1}{\lVert z\rVert_q}\Big| z\in\mathrm{ker} (A)\setminus \{0\}\right\}$ for 20 random realizations of the oversampled discrete cosine transform (DCT) matrices $A$ with $F=2$ and $F=5$ (shown as the red star-points), and compute $\inf\limits_{z\in\mathbb{R}^N} \left\{ \frac{\lVert z\rVert_1}{\lVert z\rVert_q}\Big| Az=Ax\right\}$ for each fixed $A$ with $x$ processing different sparsity levels $s\in\{2,4,6,\cdots,24\}$ (shown via the box plot). As expected, most of the ratios subject to $Az=Ax$ are upper bounded by the ones of $Az=0$. \\

\begin{proposition}
	For any $1<q\leq \infty$, $A\in\mathbb{R}^{m\times N}$ and $x\in\mathbb{R}^N$, it holds that \begin{align}
	\inf\limits_{z\in\mathbb{R}^N} \left\{ \frac{\lVert z\rVert_1}{\lVert z\rVert_q}\Big| Az=Ax\right\} \leq \inf \limits_{z\in\mathbb{R}^N} \left\{ \frac{\lVert z\rVert_1}{\lVert z\rVert_q}\Big| z\in\mathrm{ker} (A)\setminus \{0\}\right\}
	\end{align}
\end{proposition}

\section{Algorithms}

The ADMM type algorithms were used in \cite{rahimi2018scale,xu2020analysis} for solving the noiseless $\ell_1/\ell_2$ minimization problem. Unfortunately, they can not be generalized directly to the $\ell_1/\ell_q$ minimization. In fact, the minimization problem (\ref{norm_ratio}) belongs to the nonlinear fractional programming, which was comprehensively discussed in Chapter 4 of \cite{stancu2012fractional}, see also \cite{schaible1976minimization,schaible2004recent}. We investigate two kinds of methods for solving it, namely parametric methods and change of variable method. Note that it's straightforward to add a box constraint as done in Section 4.2 of \cite{rahimi2018scale} to alleviate the intrinsic drawback of $\ell_1/\ell_q$ model that tends to produce one large coefficient while suppressing the other non-zero entries. It is also worth noticing that \cite{boct2020extrapolated} proposed a proximal subgradient algorithm with extrapolations for solving a broad class of nonsmooth and nonconvex fractional program which covers the $\ell_1/\ell_2$ minimization problem. This algorithm might be adopted for our $\ell_1/\ell_q$ minimization problem, but in this section we propose two more specific algorithms for this nonlinear fractional programming where both the numerator and denominator are convex functions.

\subsection{Parametric methods}
To solve the fractional problem (\ref{norm_ratio}), we discuss a class of methods based on the solution of the following parametric problem denoted as $Q(\lambda)$ depending on a parameter $\lambda\in\mathbb{R}$: \begin{align}
\min\limits_{z\in\mathbb{R}^N} \lambda\lVert z\rVert_1-\lVert z\rVert_q \quad \text{subject to \quad $\lVert y-Az\rVert_2\leq \eta$}.
\end{align}

Let $F(\lambda)=\min\limits_{\{z\in\mathbb{R}^N|\lVert Az-y\rVert_2\leq \eta\}} (\lambda\lVert z\rVert_1-\lVert z\rVert_q)$ be the optimal value of the objective function of problem $Q(\lambda)$. Then we have the following statement, which establishes the relationship between the problem $Q(\lambda)$ and the problem (\ref{norm_ratio}).

\begin{theorem}
	Let $\bar{x}$ be an optimal solution of problem (\ref{norm_ratio}) and let $\bar{\lambda}=\frac{\lVert \bar{x}\rVert_q}{\lVert \bar{x}\rVert_1}$. Then we have \\
	1) $F(\lambda)<0$ if and only if $\lambda<\bar{\lambda}$. \\
	2) $F(\lambda)=0$ if and only if $\lambda=\bar{\lambda}$. \\
	3) $F(\lambda)>0$ if and only if $\lambda>\bar{\lambda}$.
\end{theorem}

\noindent
{\bf Remark.} It should be pointed out that \cite{wang2020accelerated} got the similar observations in parallel in its Proposition 1, but they only considered the noiseless $\ell_1/\ell_2$ model. Based on the relationship between $\ell_1/\ell_2$ and $\ell_1-\alpha\ell_2$ for certain $\alpha$, three numerical algorithms were presented there to minimize the noiseless $\ell_1/\ell_2$ model. The corresponding generalizations to the noisy $\ell_1/\ell_q$ model might be done, which though is out of the scope of this paper.\\

\noindent 
{\bf Proof.} We only proof the case 1), since the other cases can be treated similarly. If $F(\lambda)<0$, then there at least exists one solution $x'$ of $Q(\lambda)$ such that $\lambda\lVert x'\rVert_1-\lVert x'\rVert_q<0$, i.e., $\frac{\lVert x'\rVert_q}{\lVert x'\rVert_1}>\lambda$. Moreover, $x'$ is also a feasible solution of problem (\ref{norm_ratio}), hence $\frac{\lVert \bar{x}\rVert_1}{\lVert\bar{x}\rVert_q}\leq \frac{\lVert x'\rVert_1}{\lVert x'\rVert_q}$, which leads to $\frac{\lVert \bar{x}\rVert_q}{\lVert\bar{x}\rVert_1}\geq \frac{\lVert x'\rVert_q}{\lVert x'\rVert_1}$. Therefore, we have $\lambda<\bar{\lambda}$. 

Conversely, if $\lambda<\bar{\lambda}$, then there at least exists one solution $x''$ of problem (\ref{norm_ratio}) (for example $x''=\bar{x}$) such that $\frac{\lVert x''\rVert_q}{\lVert x''\rVert_1}>\lambda$. Then, we have $F(\lambda)=\min\limits_{\{z\in\mathbb{R}^N|\lVert Az-y\rVert_2\leq \eta\}} (\lambda\lVert z\rVert_1-\lVert z\rVert_q) \leq \lambda\lVert x''\rVert_1-\lVert x''\rVert_q<0$. \\

\begin{cor}
	If $F(\lambda)=0$, then an optimal solution of problem $Q(\lambda)$ is also an optimal solution of problem (\ref{norm_ratio}). 
\end{cor}
\noindent
{\bf Proof.} If $F(\lambda)=0$, then Theorem 3 implies that $\lambda=\bar{\lambda}$. If $x'$ is an optimal solution of problem $Q(\lambda)$, then $\lambda\lVert x'\rVert_1-\lVert x'\rVert_q=0$. Hence, $\frac{\lVert x'\rVert_1}{\lVert x'\Vert_q}=\lambda=\bar{\lambda}$, that is, $x'$ is optimal solution of problem (\ref{norm_ratio}). \\

\begin{figure}[htbp]
	\centering
	\includegraphics[width=\textwidth,height=0.4\textheight]{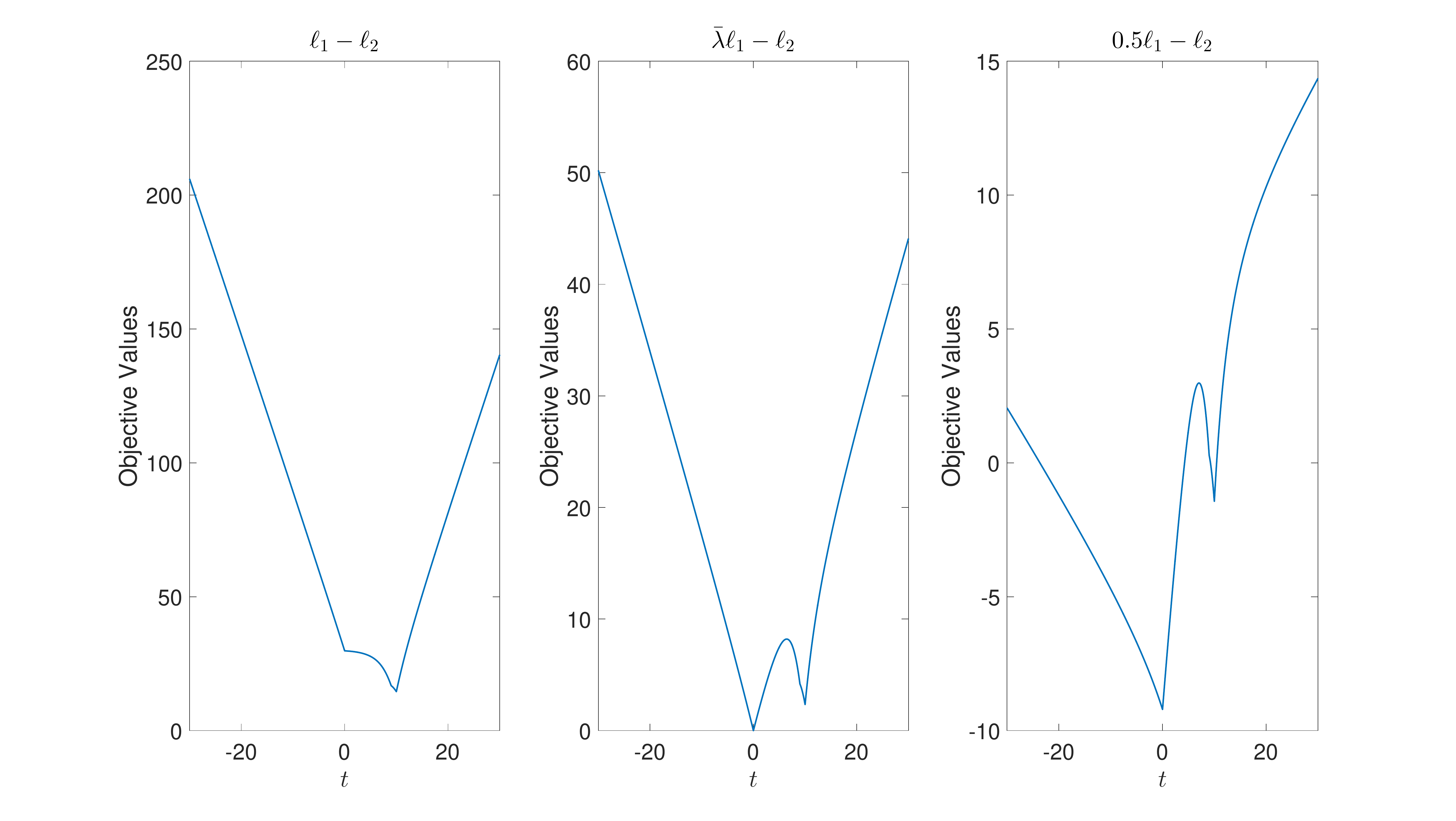}
	\caption{The objective functions of $\lambda\ell_1-\ell_2$ with $\lambda=1, \bar{\lambda}, 0$ for the toy example.} \label{fig:parametric}
\end{figure}

To illustrate the above results, we revisit the toy example discussed in Section 3 and present the objective functions of $\lambda\ell_1-\ell_2$ with $\lambda=1,\bar{\lambda},0$ in Figure \ref{fig:parametric}. Here $\bar{\lambda}=\frac{\lVert \bar{x}\rVert_2}{\lVert \bar{x}\rVert_1}$ with $\bar{x}$ being an optimal solution of problem (5) for $q=2$.  As we can see, $F(\bar{\lambda})=0$, while $F(1)>0$ and $F(0.5)<0$. As a result, finding a solution to the nonlinear fractional programming problem (\ref{norm_ratio}) is equivalent to the determination of $\lambda=\bar{\lambda}$ for which $F(\lambda)=0$, i.e., the determination of the root of the nonlinear equation $F(\lambda)=0$. It can be shown that this root is unique. The methods used for solving the equation $F(\lambda)=0$ will generate the solution algorithms for this nonlinear fractional programming problem (\ref{norm_ratio}). In fact, the function $F(\cdot)$ has many nice properties as shown in the following proposition, see Theorem 4.5.2 and its proof in \cite{stancu2012fractional} for more details.

\begin{proposition}
	The function $F(\lambda)$ is continuous, concave and strictly increasing in $\lambda\in\mathbb{R}$. 
\end{proposition}

Then, the parametric method (PM) for solving the problem (\ref{norm_ratio}) is summarized in Algorithm 1.

\begin{algorithm}[!h]
	\caption{Solving the problem (\ref{norm_ratio}) with the PM}
	\vspace*{0.5em}
	\begin{algorithmic}	
		\STATE {\textbf{Step 1}: Take $\lambda=\lambda_1$, such that $F(\lambda_1)\leq 0$.}\\
		\STATE {\textbf{Step 2}: Solve problem $Q(\lambda)$. If $|F(\lambda)|\leq \delta$, stop. Otherwise, go to Step 3.} \\
		\STATE {\textbf{Step 3}: Let $\lambda=\frac{\lVert x^{*}\rVert_q}{\lVert x^{*}\rVert_1}$, where $x^{*}$ is an optimal solution of $Q(\lambda)$ obtained in Step 2. Repeat Step 2.}
	\end{algorithmic}
\end{algorithm} 

Note that, in Step 1, $\lambda_1=0$ or $\lambda_1=\lVert \hat{x}\rVert_q/\lVert \hat{x}\rVert_1$, where $\hat{x}$ is a feasible solution of the problem. $\delta$ is a very small non-negative given constant, for instance $\delta=10^{-5}$. And we use a difference of convex functions (DCA) algorithm introduced by \cite{tao1997convex,tao1998dc} to solve $Q(\lambda)$, see Algorithm 2. 

\begin{algorithm}[!h]
	\caption{Solving the problem $Q(\lambda)$ using DCA.} 
	\vspace*{0.5em}
	\begin{algorithmic}
		\STATE {\textbf{Initialization}: Set $x^{(0)}=0$, $k=0$.}
		\STATE {\textbf{Iteration}: Repeat until $k>Max=100$ or $\lVert x^{(k+1)}-x^{(k)}\rVert_2/\max\{\lVert x^{(k)}\rVert_2,1\}<10^{-8}$, \begin{align}
			x^{(k+1)}=\begin{cases}
			\mathop{\mathrm{arg\,min}}\limits_{\{z\in\mathbb{R}^N:\lVert y-Az\rVert_2\leq \eta\}} \lambda\lVert z\rVert_1 & \text{if $x^{(k)}=0$,} \\
			\mathop{\mathrm{arg\,min}}\limits_{\{z\in\mathbb{R}^N:\lVert y-Az\rVert_2\leq \eta\}} \lambda\lVert z\rVert_1-(\lVert x^{(k)}\rVert_q^{1-q}|x^{(k)}|^{q-2}x^{(k)})^Tz & \text{otherwise.} \\
			\end{cases}
			\end{align} }
		\STATE {\textbf{Update iteration}: $k=k+1$.}
	\end{algorithmic}
\end{algorithm} 

As we can see, however, the parametric methods involve two iterations, which can inevitably result in a problem of low speed. To resolve this speed issue, we introduce a more direct and faster solver, namely the change of variable method.

\subsection{Change of variable method}

Let $v=\frac{z}{\lVert z\rVert_1}\in\mathbb{R}^N$ and $t=\frac{1}{\lVert z\rVert_1}\in\mathbb{R^{+}}$. Then, for any $1<q\leq \infty$, the minimization problem (\ref{norm_ratio}) is equivalent to the following problem: \begin{align}
\min\limits_{v\in\mathbb{R}^N, t\in\mathbb{R}^{+}} \frac{1}{\lVert v\rVert_q} \quad \text{subject to \quad $t>0$, $\lVert y-Av/t\rVert_2\leq \eta$ and $\lVert v\rVert_1=1$}.\label{cvm}
\end{align}

Now let the equality constraint $\lVert v\rVert_1=1$ be replaced by $\lVert v\rVert_1\leq 1$, and change the minimization problem to a maximization problem. Thus, it suffices to solve \begin{align}
\max\limits_{v\in\mathbb{R}^N, t\in\mathbb{R}^{+}} \lVert v\rVert_q \quad \text{subject to \quad $t\geq t_0$, $\lVert y-Av/t\rVert_2\leq \eta$ and $\lVert v\rVert_1\leq 1$}, \label{maxq}
\end{align}
where $t_0=1/a$ for some $a\geq \max\{\lVert z\rVert_1| \lVert y-Az\rVert_2\leq \eta \}$. More details concerning the reason why (\ref{cvm}) and (\ref{maxq}) have the same set of optimal solutions are referred to the arguments in Section 3 of \cite{schaible1976minimization}. By denoting the solution by $\hat{v}$ and $\hat{t}$, our final recovered signal goes to $\hat{x}=\hat{v}/\hat{t}$.

Since the problem (\ref{maxq}) is a convex-concave problem, it can be solved via a convex-concave procedure (see \cite{lipp2016variations} for details). The basic CCP algorithm goes as follows: 

\begin{algorithm}[!h]	
	\caption{CCP to solve (\ref{maxq}).}
	\vspace*{0.5em}
	\begin{algorithmic}
		\STATE {\textbf{Input}: measurement matrix $A$, measurement vector $y$, error bound $\eta$, and $t_0$.}\\
		\STATE {\textbf{Initialization}: Given an initial point $v^{0}=\frac{x^{(0)}}{\lVert x^{(0)}\rVert_1}$. Let $k=0$.}\\
		\STATE {\textbf{Iteration}: Repeat until a stopping criterion is met at $k=\bar{n}$. }\\
		\STATE {1. \textit{Convexify}. Linearize $\lVert v\rVert_q$ with the approximation $$\lVert v^{(k)}\rVert_q+\nabla (\lVert v\rVert_q)_{v=v^{(k)}}^T(v-v^{(k)})=\lVert v^{(k)}\rVert_q+\left[\lVert v^{(k)}\rVert_q^{1-q}|v^{(k)}|^{q-1}\mathrm{sign}(v^{(k)})\right]^T(v-v^{(k)}).$$}
		\vspace{-1.5em}
		\STATE {2. \textit{Solve}. Set the value of $v^{(k+1)}\in\mathbb{R}^N, t^{(k+1)}\in \mathbb{R}^{+}$ to be a solution of the linear program \begin{align}
			&\max\limits_{v\in\mathbb{R}^N, t\in\mathbb{R}^{+}}\,\lVert v^{(k)}\rVert_q+\left[\lVert v^{(k)}\rVert_q^{1-q}|v^{(k)}|^{q-1}\mathrm{sign}(v^{(k)})\right]^T(v-v^{(k)})  \nonumber \\
			&\quad\text{subject to \quad $t\geq t_0$, $\lVert y-Av/t\rVert_2\leq \eta$ and $\lVert v\rVert_1\leq 1$}. \label{ccp} \end{align} }
		\vspace*{-1.5em}
		\STATE {3. \textit{Update iteration}: $k=k+1$.} \\
		\STATE {\textbf{Output}: The recovered signal $\hat{x}=v^{(\bar{n})}/{t^{(\bar{n})}}$.}
	\end{algorithmic}
\end{algorithm}

We choose $x^{(0)}$ to be the solution of $\ell_1$-minimization problem. When $q=\infty$, and if the index to achieve the $\ell_\infty$ norm of $v^{(k)}$ is $j$, i.e., $|v^{(k)}_j|=\lVert v^{(k)}\rVert_{\infty}$, then the linearized term for $\lVert v\Vert_\infty$ at $v^{(k)}$ will be $\lVert v^{(k)}\rVert_{\infty}+\mathrm{sign}(v^{(k)}_j)(v_j-v^{(k)}_j)$. 

In addition, for the case $q=\infty$, a method based on directly solving a linear program (LP) is also presented here. Let the constraint convex set $T=\{(v\in\mathbb{R}^{N}, t\in\mathbb{R}^{+})|t\geq t_0, \lVert y-Av/t\rVert_2\leq \eta, \lVert v\rVert_1\leq 1\}$. Thus the problem (\ref{maxq}) goes to \begin{align}
\max\limits_{(v\in\mathbb{R}^{N}, t\in\mathbb{R}^{+})\in T} \lVert v\rVert_{\infty}=\max_{1\leq i\leq N} \max\limits_{(v\in\mathbb{R}^{N}, t\in\mathbb{R}^{+})\in T}  |v_i|. \label{linear}
\end{align}
Since $|v_i|=v_i^{+}+v_i^{-}$ and $v_i=v_i^{+}-v_i^{-}$, where $v_i^{+}=\max(v_i,0)$ and $v_i^{-}=\max(-v_i,0)$, (\ref{linear}) can be converted to solve $N$ linear programs: \begin{align}
\max_{1\leq i\leq N}\left\{ \max\limits_{(\tilde{v}\in\mathbb{R}^{N+1},t\in\mathbb{R}^{+})} \tilde{v}_i+\tilde{v}_{i+1} 
\text{\quad subject to \quad $t\geq t_0$, $\lVert y-\tilde{A}\tilde{v}/t\rVert_2\leq \eta$, $\lVert \tilde{v}\rVert_1\leq 1$, $\tilde{v}_i\geq 0$, $\tilde{v}_{i+1}\geq 0$} \right\},
\end{align}
where $\tilde{v}=(v_1,\cdots,v_i^{+},v_i^{-1},\cdots,v_N)^T$ and $\tilde{A}=(a_1,\cdots,a_i,-a_i,\cdots,a_N)$.

\section{Numerical experiments}

In the following experiments, we consider two types of measurement matrices, i.e., Gaussian random matrix and oversampled discrete cosine transform (DCT) matrix. Specifically, for the Gaussian random matrix, it is generated as $\frac{1}{\sqrt{m}}$ times an $m\times N$ matrix with entries drawn from i.i.d. standard normal distribution. For the oversampled DCT matrix, we use $A=[a_1,a_2,\cdots,a_N]\in\mathbb{R}^{m\times N}$ with $a_j=\frac{1}{\sqrt{m}}\cos\left(\frac{2\pi w(j-1)}{F}\right), j=1,2,\cdots,N$, and $w$ is a random vector uniformly distributed in $[0,1]^{m}$. An important property of DCT matrix is its high coherence where a larger $F$ yields a more coherent matrix.

\subsection{A test with PM and CCP}
\begin{figure}[htbp]
	\centering
	\includegraphics[width=\textwidth,height=0.4\textheight]{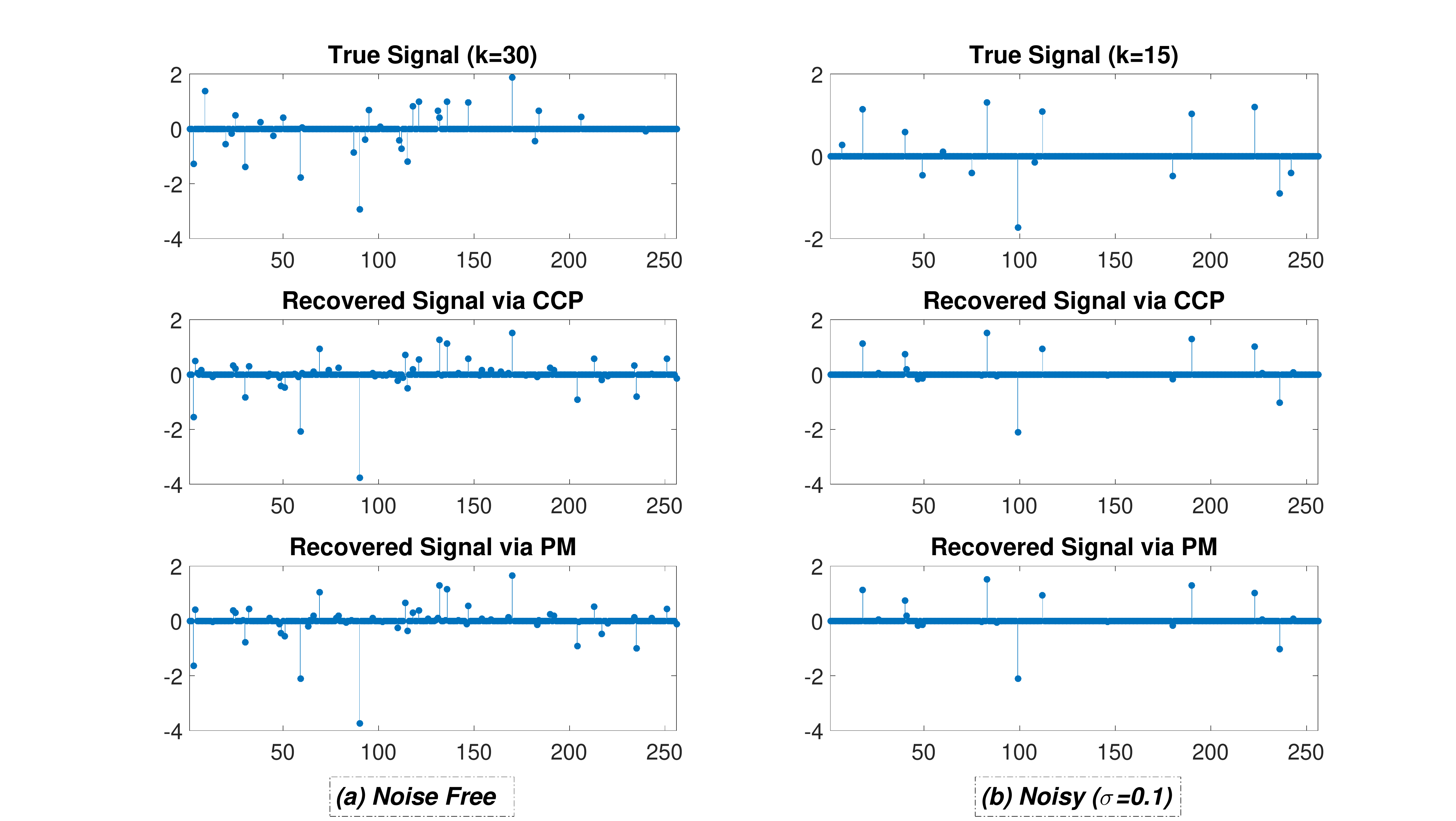} 
	\caption{Reconstruction comparison between PM and CCP for $q=2$.} \label{PM_CCP}
\end{figure}

We first compare the performances of the PM and CCP with $q=2$ for a sparse signal $x\in\mathbb{R}^{256}$ reconstruction with a Gaussian random measurement matrix $A\in\mathbb{R}^{64\times 256}$. We considered two cases, one is that the true signal $x$ has a sparsity level of 30 and the measurements are noise free, the other case is that the true signal is 15-sparse and the measurements are noisy with $\sigma=0.1$. As shown in Figure \ref{PM_CCP}, in both cases the reconstructed signals via PM and CCP methods are almost the same. Since the CCP method is much faster than PM methods, we choose to adopt the former in our numerical experiments to solve the problem (\ref{norm_ratio}).

\begin{figure}[htbp]
	\centering
	\includegraphics[width=\textwidth,height=0.4\textheight]{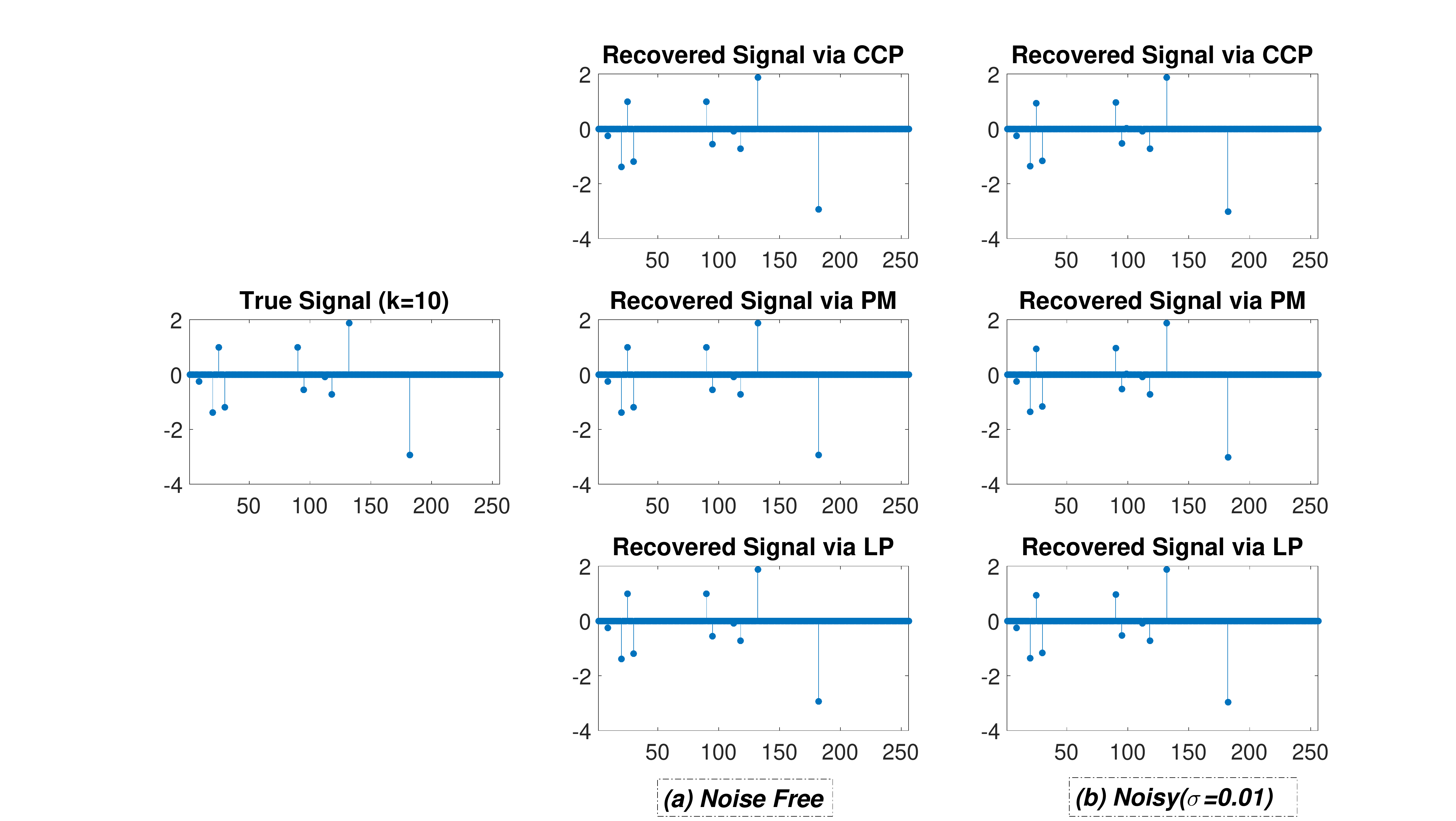} 
	\caption{Reconstruction comparison between PM, CCP and LP for $q=\infty$.} \label{infinity_comparison}
\end{figure}

Moreover, for the case of $q=\infty$, we compared three different algorithms, i.e., PM, CCP and LP, to recover a 10-sparse signal without noise and with noise ($\sigma=0.01$) in Figure \ref{infinity_comparison}. The recovered results for all these three algorithms are almost the same. When there is no noise in the measurements, a perfect recovery can be achieved via $\ell_1/\ell_\infty$, while it would not be impaired almost at all with slightly noisy measurements. 

\subsection{Choice of parameter $q$}
In order to understand what role the parameter $q$ plays in recovering sparse signals, we carried out a simulation study for $\ell_1/\ell_q$ with different parameters $q$, varying among $\{1.1, 1.5, 2, 2.5, 5,\infty\}$. In this study, $A$ is a $64\times 256$ random matrix generated as Gaussian. The true signal $x\in\mathbb{R}^{256}$ is simulated as $k$-sparse with $k$ in the set $\{6, 8,10,12,\cdots,32\}$. The support of $x$ is a random index set and its non-zero entries follow a standard normal distribution. For each $q$, we replicated the experiments 100 times with different $A$ and $x$. It's recorded as one success if the relative error $\frac{\lVert \hat{x}-x\rVert_2}{\lVert x\rVert_2}\leq 10^{-3}$. 

\begin{figure}[htbp]
	\centering
	\includegraphics[width=\textwidth,height=0.4\textheight]{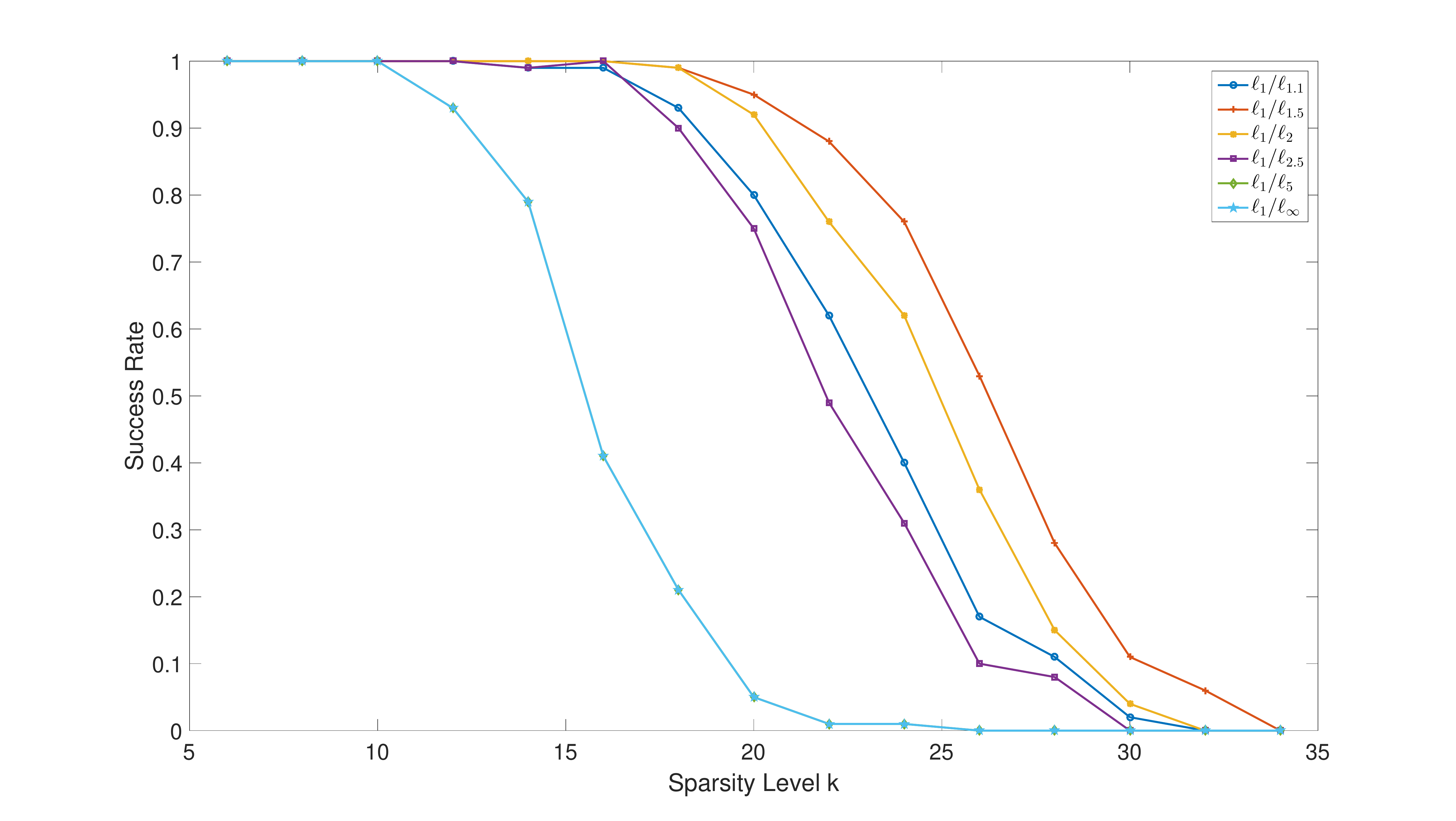}
	\caption{Recovery performance comparison for Gaussian random matrices with different $q$ for exactly sparse signals.} \label{differentq}
\end{figure}

In Figure \ref{differentq}, it shows the success rate over the 100 replicates for various values of parameter $q$ and sparsity level $k$. From this figure, we see that $q=1.5$ is the best among all tested values of $q$. And the results for $q=1.1$ and $q=2$ are better than those for $q=5$ and $q=\infty$. Basically, when $q$ is too closer to 1, the objective function is more non-convex and thus more difficult to solve. On the other hand, if $q$ is too large, the iterations are more likely to stop at a local minima far from the global one. Hence, in the following comparison study, we set $q=1.5$. 

\subsection{Comparison on different recovery methods}

\begin{figure}[htbp]
	\centering
	\includegraphics[width=\textwidth,height=0.4\textheight]{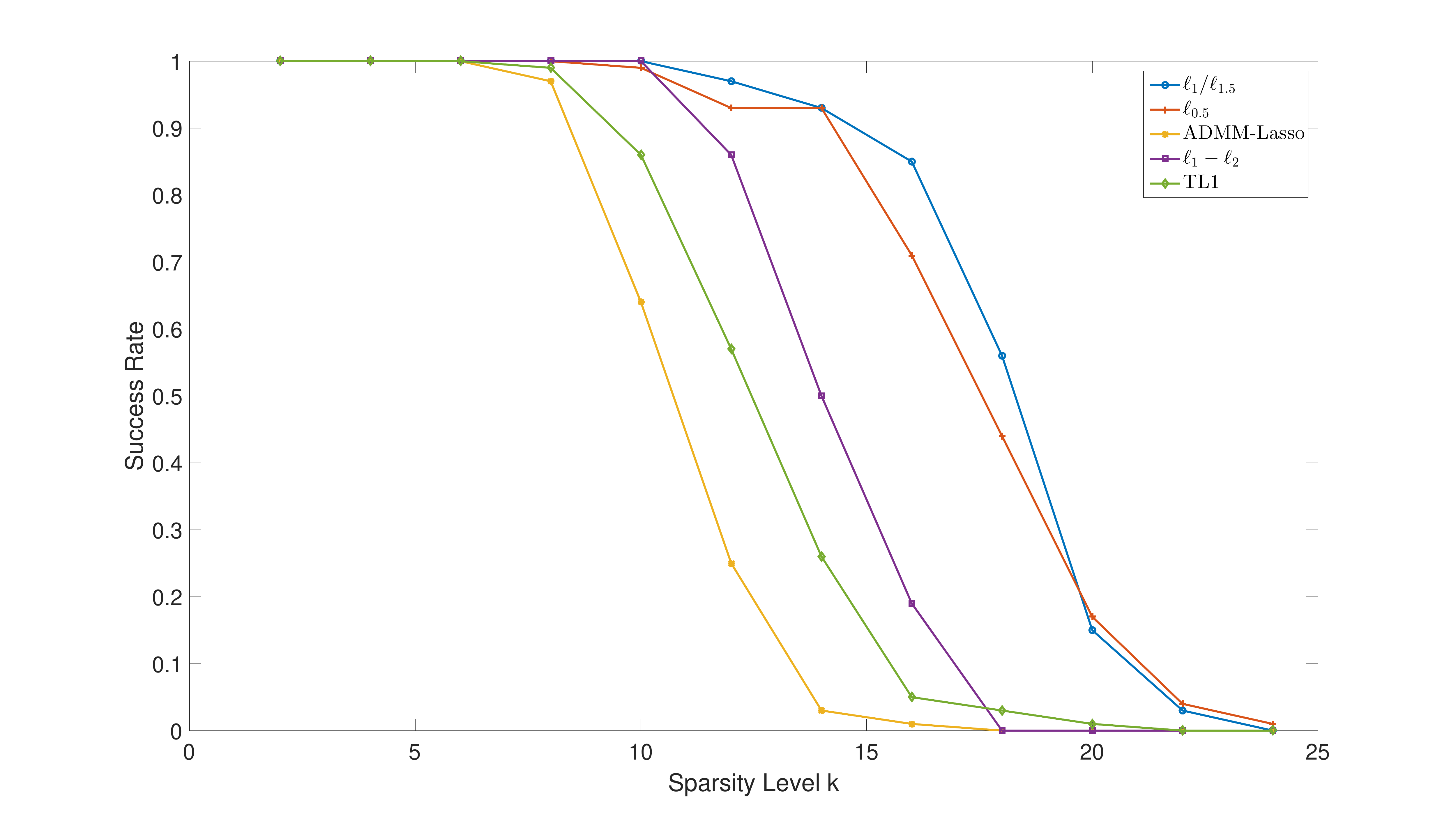}
	\caption{Recovery performance comparison for different algorithms for Gaussian random matrices.} \label{fig:4}
\end{figure}

\begin{figure}[htbp]
	\centering
	\begin{minipage}[t]{0.48\textwidth}
		\centering
		\includegraphics[width=3.3in,height=0.36\textheight]{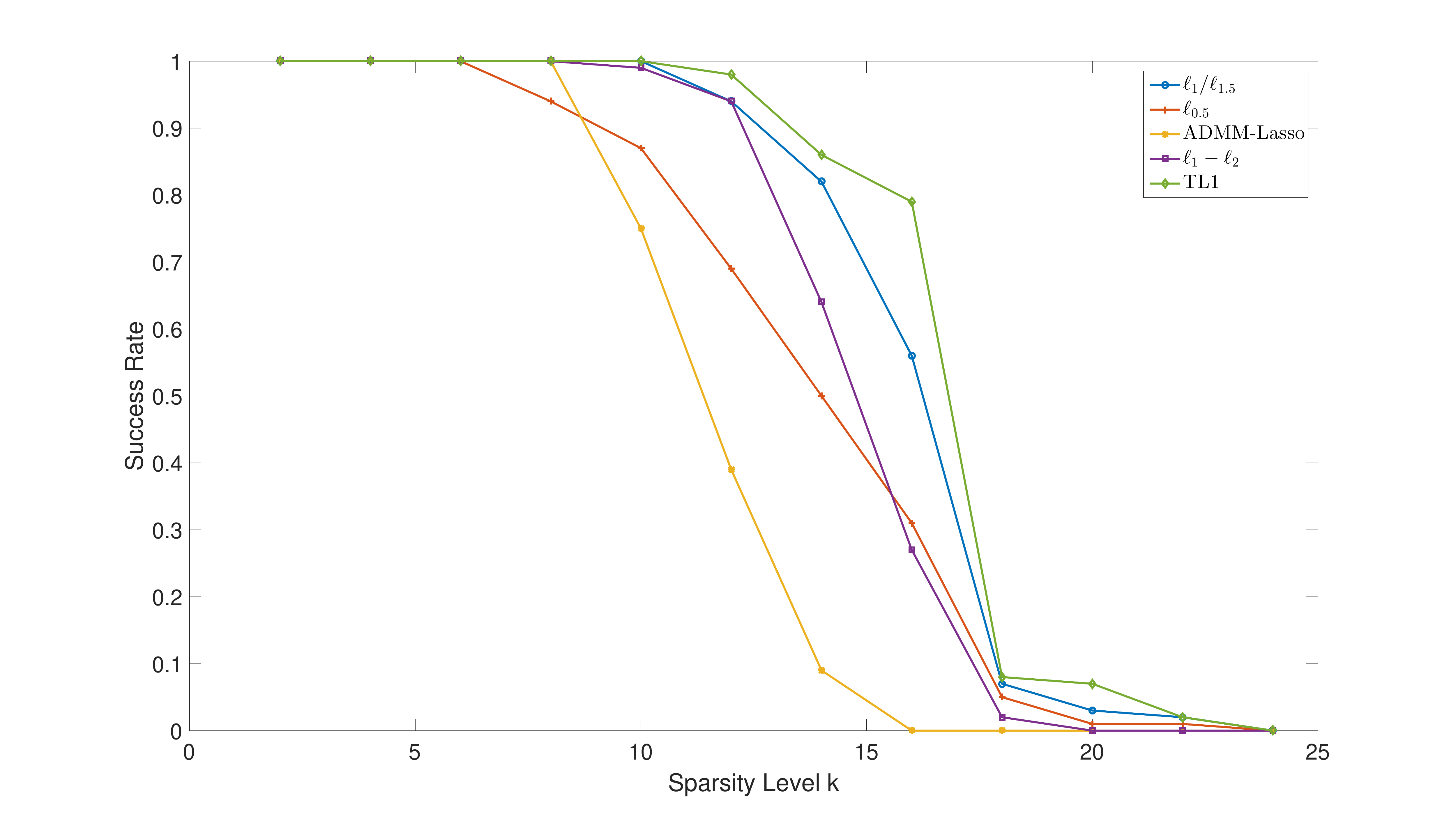}
	\end{minipage}
	\begin{minipage}[t]{0.48\textwidth}
		\centering
		\includegraphics[width=3.3in,height=0.36\textheight]{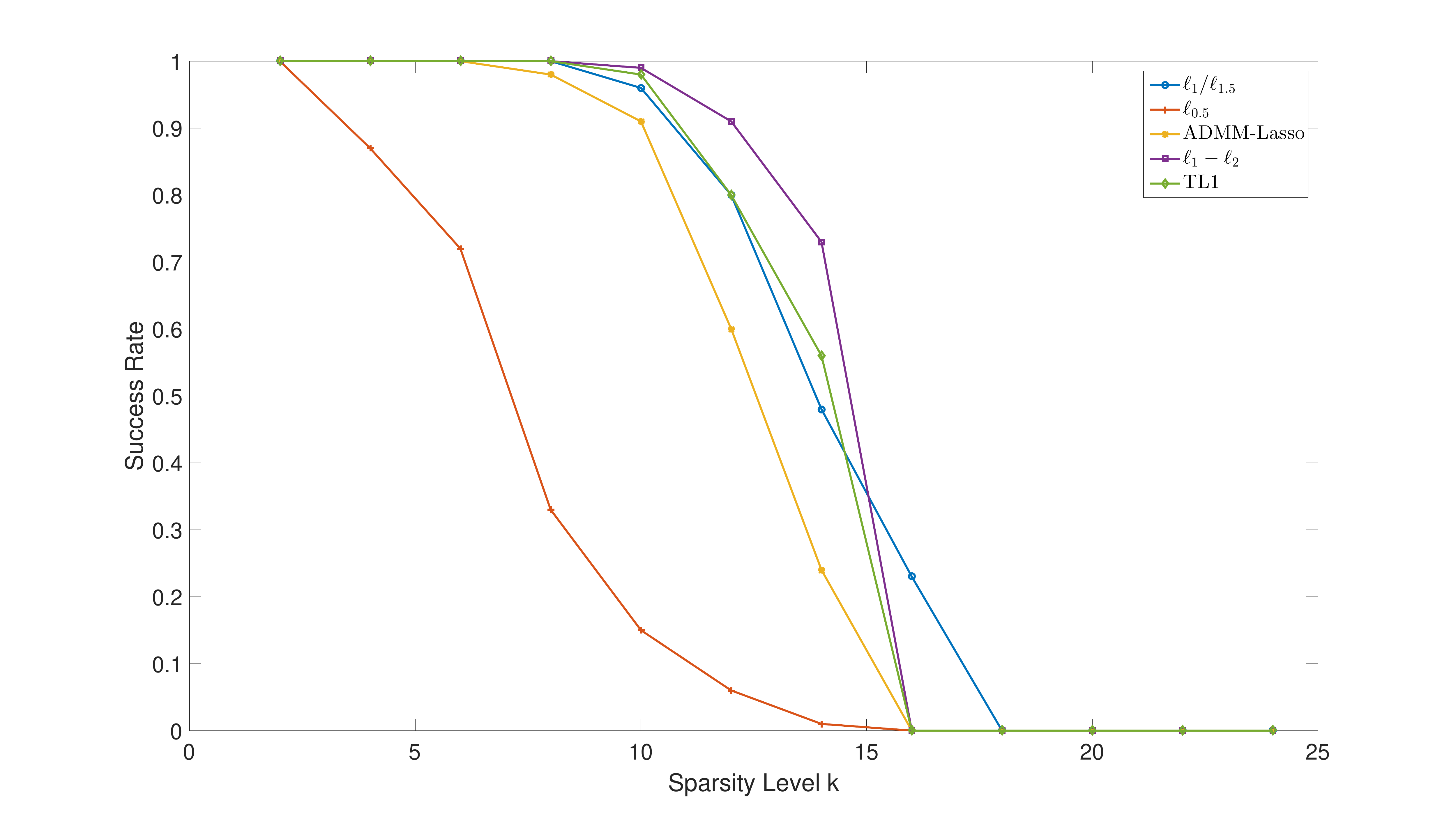}
	\end{minipage}
\caption{Recovery performance comparison for different algorithms for DCT random matrices with $F=5$ (Left Panel) and $F=10$ (Right Panel).} \label{fig:5}
\end{figure}

In this subsection, we compare the proposed $\ell_1/\ell_{1.5}$ with other state-of-the-art sparse signal recovery methods including  ADMM-Lasso, $\ell_{0.5}$, $\ell_1-\ell_2$ and TL1.  In all these comparisons, we fix the size of the measurement matrices as $64\times 1024$. The true signal $x\in\mathbb{R}^{1024}$ is simulated as $k$-sparse with $k$ in the set $\{2,4,6,8,\cdots,24\}$. The support of $x$ is a random index set and its non-zero entries follow a standard normal distribution. For each recovery method, we replicated the experiments 100 times with different $A$ and $x$ and evaluated its performance in terms of success rate for Gaussian random matrices, DCT random matrices with $F=5$ (low coherence) and $F=10$ (high coherence). The corresponding results are presented in Figure \ref{fig:4} and Figure \ref{fig:5}, respectively. 

As shown in Figure \ref{fig:4}, the $\ell_1/\ell_{1.5}$ gives the best result for the Gaussian case, even outperforms the $\ell_{0.5}$ model which is used to be the best. For the two coherent cases listed in Figure \ref{fig:5},  we can observe that although the $\ell_1/\ell_{1.5}$ is not the best, but it is still comparable to the best ones (the TL1 for $F=5$ and the $\ell_1-\ell_2$ for $F=10$). Thus, the $\ell_1/\ell_{1.5}$ model proposed in the paper gives satisfactory and robust recovery results no matter whether the measurement matrix is coherent or not. 

\section{Conclusion}

In this paper, we studied the sparse signal recovery approach via minimizing the $q$-ratio sparsity. For the case $1<q\leq \infty$, it reduces to a problem of minimizing the ratio of $\ell_1$ and $\ell_q$ norms. We gave a verifiable sufficient condition for the exact sparse recovery and established the corresponding reconstruction error bounds in terms of $q$-ratio CMSV. Two computational algorithms were proposed to approximately solve this non-convex problem. In addition, varieties of numerical experiments were conducted to illustrate our results. 

Minimization of the $q$-ratio sparsity in the case of $0<q\leq 1$ is essentially different from the case of $q\in(1,\infty]$ studied in the present paper. Both its theoretical analysis and computational algorithms are challenging and left for future work. Our analyses based on $q$-ratio CMSV in this paper do not give better recoverability and stability conditions compared to the ones given for $\ell_1$ minimization. How to obtain sharper sufficient conditions remains open. Other works that are also worth exploring in the future include the study on unconstrained models, and the explorations of their applications in image processing and machine learning.

\bibliographystyle{spmpsci}      

\bibliography{norm_ratio}

\begin{thebibliography}{10}
\providecommand{\url}[1]{{#1}}
\providecommand{\urlprefix}{URL }
\expandafter\ifx\csname urlstyle\endcsname\relax
  \providecommand{\doi}[1]{DOI~\discretionary{}{}{}#1}\else
  \providecommand{\doi}{DOI~\discretionary{}{}{}\begingroup
  \urlstyle{rm}\Url}\fi

\bibitem{boct2020extrapolated}
Bo{\c{t}}, R.I., Dao, M.N., Li, G.: Extrapolated proximal subgradient
  algorithms for nonconvex and nonsmooth fractional programs.
\newblock arXiv preprint arXiv:2003.04124  (2020)

\bibitem{boyd2011distributed}
Boyd, S., Parikh, N., Chu, E., Peleato, B., Eckstein, J., et~al.: Distributed
  optimization and statistical learning via the alternating direction method of
  multipliers.
\newblock Foundations and Trends{\textregistered} in Machine Learning
  \textbf{3}(1), 1--122 (2011)

\bibitem{candes2007dantzig}
Candes, E., Tao, T.: The dantzig selector: Statistical estimation when p is
  much larger than n.
\newblock Annals of Statistics \textbf{35}(6), 2313--2351 (2007)

\bibitem{chartrand2007exact}
Chartrand, R.: Exact reconstruction of sparse signals via nonconvex
  minimization.
\newblock IEEE Signal Processing Letters \textbf{14}(10), 707--710 (2007)

\bibitem{chartrand2008restricted}
Chartrand, R., Staneva, V.: Restricted isometry properties and nonconvex
  compressive sensing.
\newblock Inverse Problems \textbf{24}(3), 035020 (2008)

\bibitem{chartrand2008iteratively}
Chartrand, R., Yin, W.: Iteratively reweighted algorithms for compressive
  sensing.
\newblock In: Acoustics, speech and signal processing, 2008. ICASSP 2008. IEEE
  international conference on, pp. 3869--3872. IEEE (2008)

\bibitem{chen1998}
Chen, S.S., Donoho, D.L., Saunders, M.A.: Atomic decomposition by basis
  pursuit.
\newblock SIAM Journal on Scientific Computing \textbf{20}, 33--61 (1998)

\bibitem{cohen2009compressed}
Cohen, A., Dahmen, W., DeVore, R.: Compressed sensing and best $k$-term
  approximation.
\newblock Journal of the American Mathematical Society \textbf{22}(1), 211--231
  (2009)

\bibitem{donoho2006compressed}
Donoho, D.L.: Compressed sensing.
\newblock IEEE Transactions on Information Theory \textbf{52}(4), 1289--1306
  (2006)

\bibitem{eldar2012compressed}
Eldar, Y.C., Kutyniok, G.: Compressed sensing: theory and applications.
\newblock Cambridge University Press (2012)

\bibitem{esser2013method}
Esser, E., Lou, Y., Xin, J.: A method for finding structured sparse solutions
  to nonnegative least squares problems with applications.
\newblock SIAM Journal on Imaging Sciences \textbf{6}(4), 2010--2046 (2013)

\bibitem{fan2001variable}
Fan, J., Li, R.: Variable selection via nonconcave penalized likelihood and its
  oracle properties.
\newblock Journal of the American Statistical Association \textbf{96}(456),
  1348--1360 (2001)

\bibitem{foucart2009sparsest}
Foucart, S., Lai, M.J.: Sparsest solutions of underdetermined linear systems
  via $\ell_q$-minimization for $0<q\leq 1$.
\newblock Applied and {C}omputational {H}armonic {A}nalysis \textbf{26}(3),
  395--407 (2009)

\bibitem{foucart2013mathematical}
Foucart, S., Rauhut, H.: A {M}athematical {I}ntroduction to {C}ompressive
  {S}ensing, vol.~1.
\newblock Birkh{\"a}user Basel (2013)

\bibitem{hurley2009comparing}
Hurley, N., Rickard, S.: Comparing measures of sparsity.
\newblock IEEE Transactions on Information Theory \textbf{55}(10), 4723--4741
  (2009)

\bibitem{jia2018sparse}
Jia, X., Zhao, M., Di, Y., Li, P., Lee, J.: Sparse filtering with the
  generalized lp/lq norm and its applications to the condition monitoring of
  rotating machinery.
\newblock Mechanical Systems and Signal Processing \textbf{102}, 198--213
  (2018)

\bibitem{lipp2016variations}
Lipp, T., Boyd, S.: Variations and extension of the convex--concave procedure.
\newblock Optimization and Engineering \textbf{17}(2), 263--287 (2016)

\bibitem{lopes2016unknown}
Lopes, M.E.: Unknown sparsity in compressed sensing: Denoising and inference.
\newblock IEEE Transactions on Information Theory \textbf{62}(9), 5145--5166
  (2016)

\bibitem{lou2018fast}
Lou, Y., Yan, M.: Fast {L}1--{L}2 minimization via a proximal operator.
\newblock Journal of Scientific Computing \textbf{74}(2), 767--785 (2018)

\bibitem{lou2015computing}
Lou, Y., Yin, P., He, Q., Xin, J.: Computing sparse representation in a highly
  coherent dictionary based on difference of {L}1 and {L}2.
\newblock Journal of Scientific Computing \textbf{64}(1), 178--196 (2015)

\bibitem{natarajan1995sparse}
Natarajan, B.K.: Sparse approximate solutions to linear systems.
\newblock SIAM journal on computing \textbf{24}(2), 227--234 (1995)

\bibitem{pastor2015mathematics}
Pastor, G., Mora-Jim{\'e}nez, I., J{\"a}ntti, R., Caamano, A.J.: Mathematics of
  sparsity and entropy: Axioms core functions and sparse recovery.
\newblock arXiv preprint arXiv:1501.05126  (2015)

\bibitem{petrosyan2019reconstruction}
Petrosyan, A., Tran, H., Webster, C.: Reconstruction of jointly sparse vectors
  via manifold optimization.
\newblock Applied Numerical Mathematics \textbf{144}, 140--150 (2019)

\bibitem{plan2013one}
Plan, Y., Vershynin, R.: One-bit compressed sensing by linear programming.
\newblock Communications on Pure and Applied Mathematics \textbf{66}(8),
  1275--1297 (2013)

\bibitem{rahimi2018scale}
Rahimi, Y., Wang, C., Dong, H., Lou, Y.: A scale-invariant approach for sparse
  signal recovery.
\newblock SIAM Journal on Scientific Computing \textbf{41}(6), A3649--A3672
  (2019)

\bibitem{schaible1976minimization}
Schaible, S.: Minimization of ratios.
\newblock Journal of Optimization Theory and Applications \textbf{19}(2),
  347--352 (1976)

\bibitem{schaible2004recent}
Schaible, S., Shi, J.: Recent developments in fractional programming:
  single-ratio and max-min case.
\newblock Nonlinear analysis and convex analysis \textbf{493506} (2004)

\bibitem{stancu2012fractional}
Stancu-Minasian, I.M.: Fractional programming: theory, methods and
  applications, vol. 409.
\newblock Springer Science \& Business Media (2012)

\bibitem{tang2011performance}
Tang, G., Nehorai, A.: Performance analysis of sparse recovery based on
  constrained minimal singular values.
\newblock IEEE Transactions on Signal Processing \textbf{59}(12), 5734--5745
  (2011)

\bibitem{tao1997convex}
Tao, P.D., An, L.T.H.: Convex analysis approach to dc programming: {T}heory,
  algorithms and applications.
\newblock Acta {M}athematica {V}ietnamica \textbf{22}(1), 289--355 (1997)

\bibitem{tao1998dc}
Tao, P.D., An, L.T.H.: A {DC} optimization algorithm for solving the
  trust-region subproblem.
\newblock {SIAM} {J}ournal on {O}ptimization \textbf{8}(2), 476--505 (1998)

\bibitem{tibshirani1996regression}
Tibshirani, R.: Regression shrinkage and selection via the lasso.
\newblock Journal of the Royal Statistical Society: Series B (Methodological)
  \textbf{58}(1), 267--288 (1996)

\bibitem{vershynin2015estimation}
Vershynin, R.: Estimation in high dimensions: a geometric perspective.
\newblock In: Sampling theory, a renaissance, pp. 3--66. Springer (2015)

\bibitem{wang2020limited}
Wang, C., Tao, M., Nagy, J., Lou, Y.: Limited-angle ct reconstruction via the
  l1/l2 minimization.
\newblock arXiv preprint arXiv:2006.00601  (2020)

\bibitem{wang2020accelerated}
Wang, C., Yan, M., Rahimi, Y., Lou, Y.: Accelerated schemes for the $ l\_1/l\_2
  $ minimization.
\newblock IEEE Transactions on Signal Processing \textbf{68}, 2660--2669 (2020)

\bibitem{xu2020analysis}
Xu, Y., Narayan, A., Tran, H., Webster, C.: Analysis of the ratio of $\ell_1$
  and $\ell_2 $ norms in compressed sensing.
\newblock arXiv preprint arXiv:2004.05873  (2020)

\bibitem{xu2012}
Xu, Z., Chang, X., Xu, F., Zhang, H.: $ {L}_{1/2} $ regularization: A
  thresholding representation theory and a fast solver.
\newblock IEEE Transactions on {N}eural {N}etworks and {L}earning {S}ystems
  \textbf{23}(7), 1013--1027 (2012)

\bibitem{yin2014ratio}
Yin, P., Esser, E., Xin, J.: Ratio and difference of $l_1$ and $l_2$ norms and
  sparse representation with coherent dictionaries.
\newblock Communications in Information and Systems \textbf{14}(2), 87--109
  (2014)

\bibitem{yin2015minimization}
Yin, P., Lou, Y., He, Q., Xin, J.: Minimization of $\ell_{1-2}$ for compressed
  sensing.
\newblock SIAM Journal on Scientific Computing \textbf{37}(1), A536--A563
  (2015)

\bibitem{zhang2010nearly}
Zhang, C.H., et~al.: Nearly unbiased variable selection under minimax concave
  penalty.
\newblock Annals of Statistics \textbf{38}(2), 894--942 (2010)

\bibitem{zc}
Zhang, H., Cheng, L.: On the constrained minimal singular values for sparse
  signal recovery.
\newblock IEEE Signal Processing Letters \textbf{19}(8), 499--502 (2012)

\bibitem{zhang2018minimization}
Zhang, S., Xin, J.: Minimization of transformed ${L}_1$ penalty: theory,
  difference of convex function algorithm, and robust application in compressed
  sensing.
\newblock Mathematical Programming \textbf{169}(1), 307--336 (2018)

\bibitem{zhou2019new}
Zhou, Z., Yu, J.: A new nonconvex sparse recovery method for compressive
  sensing.
\newblock Frontiers in Applied Mathematics and Statistics \textbf{5}, 14 (2019)

\bibitem{zhou2018q}
Zhou, Z., Yu, J.: On $ q $-ratio {CMSV} for sparse recovery.
\newblock Signal Processing \textbf{165}, 128--132 (2019)

\bibitem{zhou2018sparse}
Zhou, Z., Yu, J.: Sparse recovery based on $q$-ratio constrained minimal
  singular values.
\newblock Signal Processing \textbf{155}, 247--258 (2019)

\end{thebibliography}
%
%

\end{document}